\documentclass[journal]{IEEEtran}
\IEEEoverridecommandlockouts

\usepackage{cite}
\usepackage{algorithmic}
\usepackage[linesnumbered,ruled,vlined]{algorithm2e}
\algsetup{linenosize=\small}
\usepackage{graphicx}
\usepackage{multirow}
\usepackage{textcomp}
\usepackage{xcolor}
\usepackage{scalerel}
\usepackage{graphics}
\usepackage[utf8]{inputenc}
\usepackage{amsfonts}
\usepackage[fleqn]{amsmath}
\usepackage{tikz}
\usepackage{tablefootnote}
\usepackage{threeparttable}
\usepackage[hidelinks]{hyperref}
\usepackage{array}
\usepackage{multirow}
\usepackage{booktabs}
\usepackage{color}
\usepackage{array}
\usepackage{hhline}
\usepackage{soul}
\usepackage{comment}
\usepackage{tablefootnote}
\usepackage{url}
\usepackage{stfloats}

\usepackage{microtype} 

\newcommand*\circled[1]{\tikz[baseline=(char.base)]{\node[shape=circle,draw,inner sep=1.5pt] (char) {#1};}}

\usetikzlibrary{svg.path}

\definecolor{orcidlogocol}{HTML}{A6CE39}
\tikzset{
    orcidlogo/.pic={
        \fill[orcidlogocol] svg{M256,128c0,70.7-57.3,128-128,128C57.3,256,0,198.7,0,128C0,57.3,57.3,0,128,0C198.7,0,256,57.3,256,128z};
        \fill[white] svg{M86.3,186.2H70.9V79.1h15.4v48.4V186.2z}
        svg{M108.9,79.1h41.6c39.6,0,57,28.3,57,53.6c0,27.5-21.5,53.6-56.8,53.6h-41.8V79.1z M124.3,172.4h24.5c34.9,0,42.9-26.5,42.9-39.7c0-21.5-13.7-39.7-43.7-39.7h-23.7V172.4z}
        svg{M88.7,56.8c0,5.5-4.5,10.1-10.1,10.1c-5.6,0-10.1-4.6-10.1-10.1c0-5.6,4.5-10.1,10.1-10.1C84.2,46.7,88.7,51.3,88.7,56.8z};
    }
}

\newcommand\orcidicon[1]{\href{https://orcid.org/#1}{\mbox{\scalerel*{
                \begin{tikzpicture}[yscale=-1,transform shape]
                \pic{orcidlogo};
                \end{tikzpicture}
            }{|}}}}

\def\BibTeX{{\rm B\kern-.05em{\sc i\kern-.025em b}\kern-.08em
    T\kern-.1667em\lower.7ex\hbox{E}\kern-.125emX}}

\begin{document}
\bstctlcite{IEEEexample:BSTcontrol}
\title{A Security-aware and LUT-based CAD Flow for the Physical Synthesis of eASICs}

\author{Zain~Ul~Abideen\textsuperscript{\orcidicon{0000-0002-8865-9402}}, \IEEEmembership{Graduate Student Member,~IEEE,}
        Tiago~Diadami~Perez\textsuperscript{\orcidicon{0000-0001-6006-1938}}, \IEEEmembership{Graduate Student Member,~IEEE,}\\
       Mayler~Martins\textsuperscript{\orcidicon{0000-0002-2848-2190}},  Samuel~Pagliarini\textsuperscript{\orcidicon{0000-0002-5294-0606}}, \IEEEmembership{Member,~IEEE}
  \thanks{This work has been partially conducted in the project ``ICT programme'' which was supported by the European Union through the ESF.}
  \thanks{Z. U. Abideen, T. D. Perez, and S. Pagliarini are with the Department of Computer Systems, Centre for Hardware Security, Tallinn University of Technology (TalTech), 12616, Tallinn, Estonia (e-mail: zain.abideen@taltech.ee; tiago.perez@taltech.ee; samuel.pagliarini@taltech.ee).}
  \thanks{M. Martins is with Synopsys Inc., 690 E Middlefield Rd, Mountain View, CA  (e-mail: Mayler.Martins@synopsys.com).}
\vspace*{-1.0cm}}

\markboth{
}{Zain~Ul \MakeLowercase{\textit{et al.}}: A Security-aware and LUT-based CAD Flow for the Physical Synthesis of eASICs}

\maketitle

\begin{abstract}
Numerous threats are associated with the globalized integrated circuit (IC) supply chain, such as piracy, reverse engineering, overproduction, and malicious logic insertion. Many obfuscation approaches have been proposed to mitigate these threats by preventing an adversary from fully understanding the IC (or parts of it). The use of reconfigurable elements inside an IC is a known obfuscation technique, either as a coarse grain reconfigurable block (i.e., eFPGA) or as a fine grain element (i.e., FPGA-like look-up tables).  
This paper presents a security-aware CAD flow that is LUT-based yet still compatible with the standard cell based physical synthesis flow. More precisely, our CAD flow explores the FPGA-ASIC design space and produces heavily obfuscated designs where only small portions of the logic resemble an ASIC. Therefore, we term this specialized solution an ``embedded ASIC'' (eASIC). Nevertheless, even for heavily LUT-dominated designs, our proposed decomposition and pin swapping algorithms allow for performance gains that enable performance levels that only ASICs would otherwise achieve. On the security side, we have developed novel template-based attacks and also applied existing attacks, both oracle-free and oracle-based. Our security analysis revealed that the obfuscation rate for an SHA-256 study case should be at least 45\% for withstanding traditional attacks and at least 80\% for withstanding template-based attacks. When the 80\% obfuscated SHA-256 design is physically implemented, it achieves a remarkable frequency of 368MHz in a 65nm commercial technology, whereas its FPGA implementation (in a superior technology) achieves only 77MHz.

\end{abstract}

\begin{IEEEkeywords}
Hardware Obfuscation, Secure ASIC Design, LUT-based obfuscation, Reverse engineering, embedded ASIC
\end{IEEEkeywords}

\vspace{-2mm}
\section{Introduction} \label{sec:intro}

\IEEEPARstart{N}{owadays}, high-performance and energy-efficient integrated circuits (ICs) are enablers in a variety of application domains. However, this demands the fabrication of ICs in advanced technology nodes. Current predictions are that the sales of semiconductor devices will rise to \$680.6B in 2022, the first time this mark has been surpassed in a calendar year since 2020 \cite{costDevices}. In tandem, the majority of IC design houses are adhering to a globalized supply chain to outsource fabrication from pure-play foundries. Even very large semiconductor companies rely on the so-called fab-for-hire model \cite{Apple3nm, intelTsmc}, a framework that originates from the technological and financial challenges of developing and maintaining a foundry. The estimated cost to build a 3nm production line is \$15-20B \cite{Cost3nm}. The trend is clear: more than ever, fabless design companies rely on outsourcing the manufacturing of their ICs.

While this business model enables design houses to have access to high-end manufacturing, the integrity and trustworthiness of the ICs are potentially affected. For manufacturing an IC, the design house must share a blueprint of the IC with the foundry. This blueprint inevitably exposes all aspects of the IC and its many parts. A rogue element within the foundry can entirely or partially copy the design, i.e., the foundry and its employees are considered \emph{potential adversaries}. Many potential threats are associated with the untrusted fabrication aspect of a globalized IC supply chain \cite{Rostami2014}. Such threats include tampering, counterfeiting, reverse engineering, and overproduction.

Numerous techniques have been devised to protect against the aforementioned security threats.  Countermeasures to secure an IC also apply to a malicious end-user that can be interested in reverse engineering a design. Noteworthy examples of countermeasures are Logic Locking \cite{logic_1, logic_2, logic_3, logic_4, logic_5}, IC Camouflaging \cite{cam_1, cam_2, cam_3}, Split Manufacturing \cite{split_1, split_2}, and FPGA-like obfuscation approaches \cite{reconfigure_3, e-FPGA1, e-FPGA2, e-FPGA4, lut1, lut2, lut3, lut4, lut5}. The latter style of obfuscation attempts to exploit an FPGA (or FPGA-like) fabric, where the functionality of the circuit is hidden by the configuration and the \emph{bitstream serves as a key to unlock the design}. 

\begin{figure}[tb]
\centering \footnotesize
\includegraphics[width=1.0\linewidth]{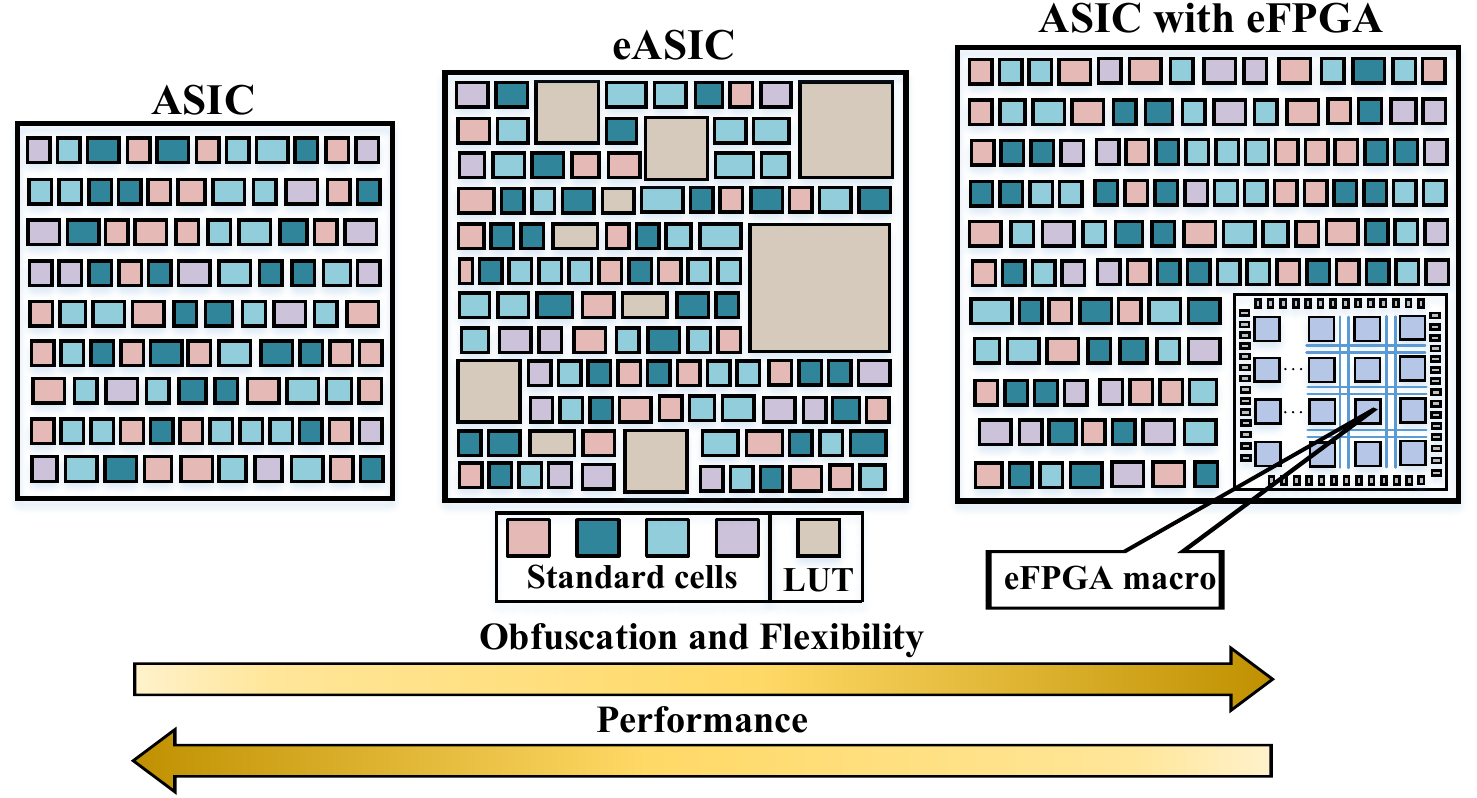}
\caption{The design obfuscation landscape.}
\centering
\label{fig:motivation}
\vspace{-2mm}
\end{figure}

Generally, the fabric in an FPGA device contains many reconfigurable blocks that can be leveraged for obfuscation purposes. The ability to reconfigure a device does incur performance penalties (i.e., FPGA vs. ASIC). Being so, custom solutions where only a small portion of the design is reconfigurable have been sought, a solution typically termed eFPGA.
This work also takes advantage of this possibility. A visualization of the obfuscation landscape is given in Fig.~\ref{fig:motivation}. As illustrated, performance increases if we move from right to left. Contrarily, obfuscation and flexibility increase if we move from left to right. However, we argue that \emph{neither extremes of the landscape are a good design point} for circuits with stringent security and performance constraints. A midpoint solution is a better trade-off, which is precisely the motivation for our work. We term our midpoint solution an ``embedded ASIC'' (\textbf{eASIC}).

In \cite{eASIC}, we have described an initial attempt at explore and automate the design spaces captured in Fig.~\ref{fig:motivation}. In this work, we extend and improve our results considerably while keeping the same general theme: we seek to obfuscate a circuit by generating a hybrid design that consists of a reconfigurable portion and static logic. The \textit{reconfigurable} part provides the obfuscation while the \textit{static} logic provides performance benefits. We perform our design space exploration at block level. Finally, the architecture of the generated block is a mix of \textit{reconfigurable} and \textit{static} cells. The reconfigurable part is implemented with programmable LUTs; the circuit is largely non-functional until it is programmed.

Earlier obfuscation techniques utilizing reconfigurable elements have focused on keeping the reconfigurable part as small as possible. Understandably, the goal would be to avoid large performance and area overheads. However, we emphasize (and later provide results) that proper hiding of the circuit's intent requires a \emph{high degree of obfuscation} that is generally not investigated in the state of the art. Consequently, the main contributions of this work are:

\begin{itemize}

    \item A CAD flow and a tool for automatically obfuscating a design, thus generating a specialized solution called eASIC that is compatible with standard-cell based flows and current design and fabrication practices.
    \item Specialized algorithms for performance improvement of eASICs, including LUT decomposition and pin swapping approaches.
    \item An analysis of performance versus obfuscation and area versus obfuscation trade-offs for numerous designs, including known benchmarks.
    \item A detailed analysis (physical synthesis) of performance, power, and area versus obfuscation for SHA-256, including tapeout-ready layouts in a 65nm commercial technology.
   \item Thorough analysis of eASIC's security against custom attacks and known oracle-based and oracle-less attacks.
\end{itemize}

In Section \ref{sec:cad_flow}, we present the secure CAD flow and its internal architecture. In Section \ref{sec:dec_alg}, we describe the algorithms utilized to improve the LUT performance, such as LUT decomposition and pin swap approaches. In Section \ref{sec:results}, we report initial results. In Section \ref{sec:physical_implementation}, we report the physical synthesis results for SHA-256. In Section \ref{sec:security_analysis}, we perform the security analysis of eASIC while considering numerous attacks. In Section \ref{sec:discussion}, we provide discussions. In Section \ref{sec:conclusion}, we conclude the paper.

\vspace{-2mm}
\section{A CAD Flow for eASIC} \label{sec:cad_flow}
\begin{figure}[tb]
\centering \footnotesize
\includegraphics[width=1.0\linewidth]{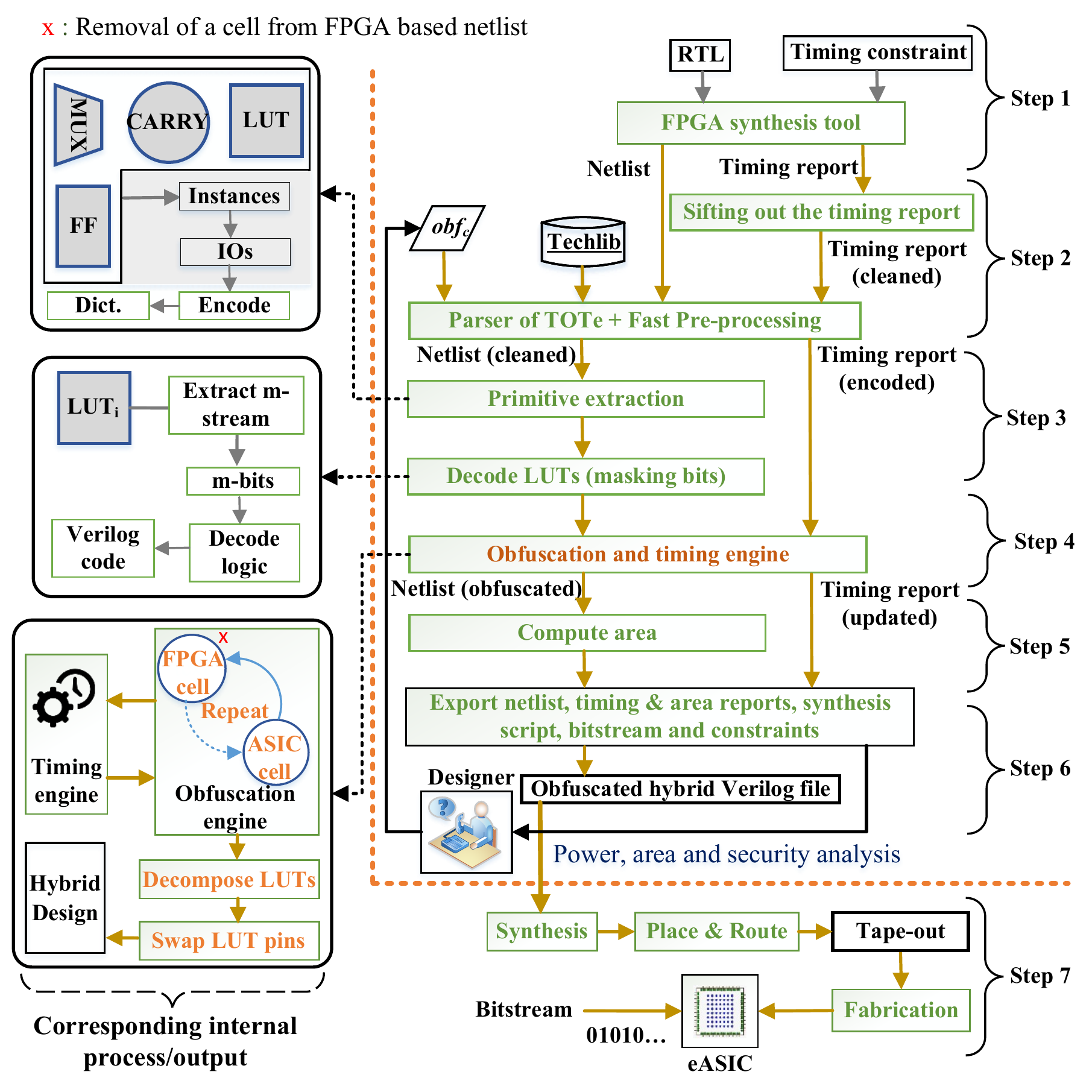} 
\caption{Overview of TOTe's obfuscation flow and its inner steps}
\centering
\label{fig:data_flow}
\vspace{-2mm}
\end{figure}

Our CAD flow utilizes a custom tool named \textbf{T}uneable Design \textbf{O}bfuscation \textbf{T}echnique using \textbf{e}ASIC (TOTe). Our custom tool produces an eASIC design with reconfigurable and static logic. The reconfigurable portion of eASIC instantiates programmable LUTs (Look Up Tables). The complete process for obfuscating a design is fully automated and infers a marginal increase in design time (when compared to a traditional ASIC flow). The complete design flow for obfuscating a design, generating an eASIC along with logical \& physical synthesis, is illustrated in Fig. \ref{fig:data_flow} and comprises a total of 8 steps, which we represent as circled numbers in the text that follows. 

In Step \circled{1}, the design under obfuscation (DUO) in register-transfer level (RTL) form is synthesized using a commercial FPGA synthesis tool. The DUO requires no special annotations, synthesis pragmas, or any other change in its representation. Outputs from Step \circled{1} are in the form of a synthesized netlist and a timing report. The netlist comprises all the typical FPGA primitives, i.e., MUXs, LUTs, and FFs. We note that at this point, the logic of the design is 100\% obfuscated since it is captured by LUTs. In very short words, the next steps of TOTe will find LUTs that are good candidates for being replaced by static logic. This is the \textbf{core functionality} of TOTe, and is illustrated on the bottom left corner of Fig.~\ref{fig:data_flow}.

Next, in Step \circled{2}, pre-processing takes place. This step aims to filter and interpret the timing report and Verilog netlist. The parsing of the timing report is a relatively trivial task. The timing report contains information that should be discarded (empty lines, headers, etc.) for which a bash script has been written. Furthermore, there are also elements that appear in the netlist that have no real purpose in an eASIC implementation, such as the buffers for IOs (i.e., IBUF and OBUF cells in Xilinx nomenclature) and for the interconnect. After removing the buffers, every analyzed path may now contain four FPGA primitives: $FF$, $CARRY$, $LUT\textsubscript{i}$, and $MUX$. TOTe encodes (hashes) the instance names to avoid lengthy string representations. The pre-processing step ends when TOTe produces a list of timed paths, where each path contains a list of hashed instances and associated delay values. 


Note that an instance can appear in many paths and also can appear in many paths under different timing arcs. Finally, the list of timed paths is sorted in ascending order. As a result, the path that has the highest delay (critical path) is referred to as $CP$ and the sum of all $CPs$ is referred to as $sumCP$\footnote{CP and SumCP are analogous to WNS and TNS in traditional static timing analysis (STA), except all paths here are assumed to pass timing checks. For simplicity, no negative values are therefore considered in this analysis.}.



Step \circled{3} is the process of primitive extraction and LUT decoding. TOTe builds a graph representation of the netlist to keep track of port connections such that the circuit structure can be preserved once optimizations are applied. Under the graph representation, primitive types are annotated for every instance; For LUTs, in particular, the tool also annotates their masking patterns (i.e., configuration bits of an individual LUT). In practice, TOTe is able to interpret the LUT encoding scheme utilized in the netlist coming from FPGA synthesis. For the case of a LUT\textsubscript{6}, the 64-bit masking pattern extracted from the netlist is converted to a truth table with 6 inputs and 1 output. The masking pattern determines which combinations of inputs generate outputs as 1s and 0s. The process is identical for smaller LUTs, which then have smaller truth tables. By using truth tables populated by the masking patterns, TOTe builds combinational logic that is equivalent to the LUT's logic. The truth tables are exported as synthesizable Verilog code. Other primitives, such as $FF$ and $MUX$, require no decoding and are directly translated to their ASIC equivalents.

TOTe comes with obfuscation and timing engines that drive the security vs. performance objectives of the tool. These engines are utilized in Step \circled{4} and are responsible for different important tasks, including timing analysis, critical path identification, and replacement of reconfigurable cells for static cells. Algorithm~\ref{alg:obfuscation_engine} describes the different operations inside the obfuscation main loop of the tool, where $L$ is a list of LUTs, $P$ is a list of timed paths, and $obf_c$ is the obfuscation criterion. The internal variables $L\textsubscript{ST}$ and $L\textsubscript{RE}$ are lists of LUTs in static and reconfigurable form, respectively. Initially, all LUTs are considered (line 2). Then, the obfuscation engine executes until the desired number of LUTs is made static (line 3), where the SIZE\_OF function returns the size of a list. Inside the obfuscation inner loop, the critical path is identified (line 4) using the FIND\_CRITICAL function, then the slowest LUT on that path is identified using the FIND\_SLOWEST function (line 5). If the identified LUT is a reconfigurable LUT (line 6), the lists of LUTs are updated (lines 7-8) and the timing engine recalculates the affected paths (line 9). If the identified LUT is not reconfigurable (line 10), the path is removed (line 11) and the loop continues (line 3). The INSERT and REMOVE functions update the lists as hinted by their names.

\begin{algorithm}[tb]
\small
\caption{TOTe's obfuscation procedure}
\label{alg:obfuscation_engine}
  \SetKwInOut{Require}{Require}
  \SetKwInOut{Ensure}{Ensure}
  \SetKwInput{KwInput}{Input}
  \SetKwInput{KwOutput}{Output}
\DontPrintSemicolon
  \KwInput{$L$ $(list$ $of$ $LUTs$), $P$ $(list$ $of$ $paths)$,  $obf_c$ $(obfuscation$ $criterion)$}
  \KwOutput{$eASIC \gets f(input)$}
  
  $L\textsubscript{ST} \gets \phi$,   $L\textsubscript{RE} \gets L$
  
  \While{SIZE\_OF {($L\textsubscript{ST}$)} $\leq$ $obf_c$}
  {
  $path$ $\gets$ FIND\_CRITICAL($P$)
  
  $lut$ $\gets$ FIND\_SLOWEST($path$)
  
      \If{$lut \in L\textsubscript{RE}$}
        {
        INSERT($lut$, $L\textsubscript{ST}$)
        
        REMOVE($lut$, $L\textsubscript{RE}$)
        
        UPDATE\_TIMING($lut$, $P$)
        }
        \Else{
        REMOVE($path$, $P$)
        }
}

  \For{each $lut \in L\textsubscript{ST}$}
    {
    DECODE($lut$)
    
    }
    
      \For{each $lut \in L\textsubscript{RE}$}
    {
    
    GEN\_CASE\_0\_1($lut$)
    
    DECOMPOSE\_OPT($lut$)
    
    SWAP\_PINS($lut$)
    }

$eASIC \gets L\textsubscript{ST} \cup L\textsubscript{RE}$
\end{algorithm}

A few additional steps take place after the obfuscation criterion has been met (lines 12-17). These steps are already related to the implementation of eASIC, but we list then here for completeness. The DECODE function operates on every LUT that was assigned to be static. From Step \circled{3}, TOTe already possesses their description in Verilog as truth tables. TOTe then executes the ASIC synthesis of the truth tables to obtain netlists composed of standard cells. The function GEN\_CASE\_0\_1 generates the `force logic' to be used for timing and power analysis during physical synthesis, otherwise each LUT would be timed for its worst timing arc instead of the actual implemented timing arc when the LUT is programmed. DECOMPOSE\_OPT decomposes the larger LUTs into smaller LUTs. Due to the complexity of this operation, we dedicate an entire section to it (see Section \ref{sec:dec_alg}). SWAP\_PINS performs a final timing optimization that is an attempt to swap the LUT pins in order to improve the delay, also discussed later in Section \ref{subsec:pin_swap_approach}. Finally, the algorithm merges $L\textsubscript{ST}$ and $L\textsubscript{RE}$ to generate eASIC and returns.

In Step \circled{5}, area estimation is performed. The estimated area of the eASIC  design is calculated as $A = A_{re} + A_{st}$, where $A_{re}$ is the area of the reconfigurable part and $A_{st}$ is the area of the static part. For calculating $A_{re}$, we sum the area of the LUTs that remain reconfigurable. For calculating $A_{st}$, we sum the area of the standard cells of the static LUTs. Later, a very precise area estimation is done in a industry-strength physical synthesis tool, where congestion is properly accounted for.


Step \circled{6} mostly relates to the generation of files that describe the eASIC intent. This step exports an obfuscated hybrid Verilog file, a timing report, and an area report. A designer can repeat this procedure until he/she achieves his obfuscation (security) and performance targets.

Finally, in Step \circled{7}, the obfuscated netlist is implemented in a commercial physical synthesis tool where traditional P\&R, CTS, DRC, etc., steps are executed and the resulting tapeout database is sent to the foundry for fabrication. Once the fabricated parts are delivered, they have to be programmed for the eASIC design to be functional. The programming step requires a bitstream, same as in an FPGA design. 

\vspace{-2mm}
\section{LUT-specific Approaches to Improve QoR} \label{sec:dec_alg}
In this section, we discuss LUT-related optimizations and decisions taken in order to make an eASIC design display the high-performance characteristics of an ASIC and the obfuscation capability of an FPGA fabric. 

\vspace{-2mm}
\subsection{Custom Standard Cell Based LUTs} \label{subsec:lut_implementation}
\begin{figure}[tb]
\centering \footnotesize
\includegraphics[width=0.9\linewidth]{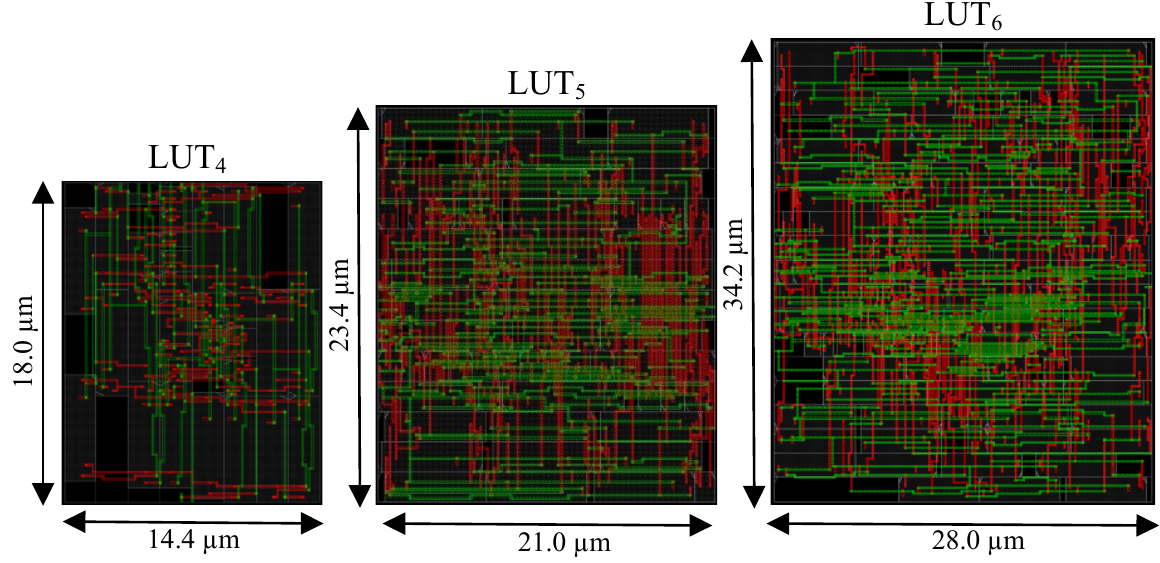}
\caption{The layout of macros for LUT\textsubscript{4}, LUT\textsubscript{5}, and LUT\textsubscript{6}. Implementation was executed in Cadence Innovus.}
\centering
\label{fig:lut_imp}
\vspace{-2mm}
\end{figure}

We have designed our own custom LUTs (LUT\textsubscript{1}, LUT\textsubscript{2},..., LUT\textsubscript{6}) out of \textbf{regular standard cells} and by following VPR's template \cite{vpr8}. The layouts for LUT\textsubscript{4}, LUT\textsubscript{5}, and LUT\textsubscript{6} macros are shown in Fig. \ref{fig:lut_imp}. Table \ref{tab:lut_blocks} shows the average delays and other characteristics of the implemented LUTs. 

We emphasize one more time that our LUTs were generated as macros composed of standard cells, thus making them compatible with standard cell based design flows. These macros are highly compact since the main design goal for them was area/density. The density achieved during physical implementation for the LUT macros is listed in Table~\ref{tab:lut_blocks}. The number of flip-flops grows with the LUT\textsubscript{i} size ($2^i$). The area for these macros approximately doubles from LUT\textsubscript{i} to LUT\textsubscript{i+1}. Compared to our previously utilized LUTs \cite{eASIC}, we have slightly increased driving strengths of the larger LUTs. 

Commercial FPGAs typically implement only one LUT size, but eASIC provides the flexibility to implement the design with different LUT sizes. This is because the generated eASIC solution is design-specific, meaning that the reconfigurability notion of an FPGA is no longer sought. Moreover, our LUT macros are highly compact, which helps placement to achieve high-density designs. Every single LUT contains a number of flip-flops for storing the configuration bits that serve as a lock for the obfuscated design. Each LUT also makes use of three extra pins ($serial\_in$, $serial\_out$, and $enable$) to configure these registers. The LUTs are serially connected to one another, forming a daisy chain that is analogous to a scan chain. The choice of a flip-flop based implementation makes our framework technology-agnostic while making the floorplan and placement almost effortless. Moreover, the LUTs themselves are also treated as regular standard cells during physical synthesis. This allows us to take full advantage of placement algorithms from commercial EDA tools, thus eliminating the need for any extra custom scripts for placing the LUT macros.


\vspace{-2mm}
\subsection{LUT Decomposition}
\begin{figure}[tb]
\centering \footnotesize
\includegraphics[width=1.0\linewidth]{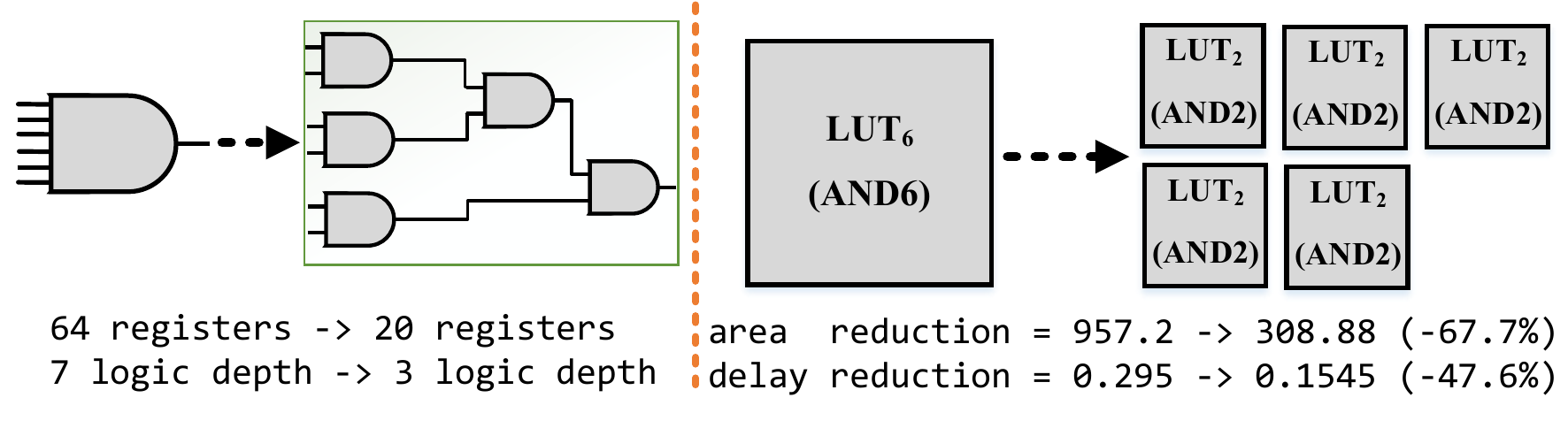}
\caption{Logic conversion and decomposition of LUT\textsubscript{6}.}
\centering
\label{fig:lut_decomp}
\vspace{-2mm}
\end{figure}

The area and delay of a LUT are directly related to its number of inputs: The area is mostly bounded by the number of sequential elements used to store the LUT's truth table, whereas the delay is proportional to the LUT's internal MUX tree. However, not all 6-input functions require a LUT\textsubscript{6} to be implemented. For instance, an AND6 can be decomposed in 5 AND2s, as presented in Fig. \ref{fig:lut_decomp}. Referring to Table \ref{tab:lut_blocks}, it is clear that the area almost doubles for each input added to the LUT while the delay increases considerably (LUT\textsubscript{6} has almost 6x more average delay than LUT\textsubscript{2}). Considering the previous example of decomposing a LUT\textsubscript{6}, the area can be reduced to less than one third of the original while the delay is reduced to approximately half of the original one. 

\begingroup
\setlength{\tabcolsep}{2.0pt} 
\renewcommand{\arraystretch}{1.1} 
\begin{table} [tb]
\footnotesize \centering
\caption{Block implementation results for $LUT\textsubscript{i}$}
\label{tab:lut_blocks}
\begin{tabular}{|p{0.8cm}|p{1.05cm}|p{1.5cm}|p{1.15cm}|p{1.84cm}|p{1.34cm}|}
\hline 
\textbf{Macro} & \textbf{Area ($\mu m^2$)} & \textbf{Density (\%)} & \textbf{Registers} & \textbf{Comb. cells} & \textbf{Avg. delay (ns)} \\ \hline
LUT\textsubscript{1} & 36.00 & 76.00 & 2 & 1 & 0.049 \\ \hline
LUT\textsubscript{2} & 64.80 & 76.26 & 4 & 1 & 0.052 \\ \hline
LUT\textsubscript{3} & 117.00 & 89.23 & 8 & 8 & 0.119 \\ \hline
LUT\textsubscript{4} & 259.20 & 85.23 & 16 & 15 & 0.192 \\ \hline
LUT\textsubscript{5} & 491.40 & 91.50 & 32 & 33 & 0.257 \\ \hline
LUT\textsubscript{6} & 957.60 & 91.09 & 64 & 36 & 0.295 \\ \hline
\end{tabular}
\vspace{-2mm}
\end{table}
\endgroup

In order to decompose our LUTs, we will make use of Functional Composition (FC) \cite{martins2012functional}, an approach that is able to perform bottom-up association of Boolean functions and control the costs as it goes. Such capability contrasts with traditional top-down functional decomposition, which does not provided a final cost until the decomposition is complete. 

\subsubsection{Functional Composition for LUTs} \label{subsec:fc_lut}

A summary of the FC paradigm and its application for LUTs will be presented. Readers can obtain more details from \cite{martins2012functional,martins2010boolean}. FC is a bottom-up paradigm that has 5 principles: 1) it uses bonded-pairs (BPs) that have a functional part (canonical implementations of a Boolean function, i.e., BDDs or truth tables) and an implementation part ( the structure that is being optimized, i.e., a fanout-free LUT circuit in this paper); 2) every BP association performs independently the functional/implementation operations, allowing for more complex implementations with simple functional operations; 3) using partial ordering and dynamic programming, all BPs with the same cost are stored together in a set (bucket), allowing the use of intermediate solutions as sub-problems and to perform associations in a cost-increasing fashion; 4) to start any FC algorithm, initial BPs are required, i.e., constants and single input variables; 5) it allows the heuristic selection of a subset of allowed functions to reduce the composition search space.

\subsubsection{Exhaustive LUT FC method} \label{subsec:fc_approach}

FC initially can be applied in an exhaustive manner, which provides fanout-free implementations that can have optimal cost. Algorithm~\ref{alg:fcopt} is able to generate all minimal LUT fanout-free implementations for functions up to four variables. The algorithm to generate functions with $N$ inputs consists of generating all implementations using LUT($N-1$) that have a smaller cost than the LUT($N$). Functions with up to 2 inputs are by definition not decomposable. For functions containing $N$ inputs, a set of functions that will serve as Boolean operators is required. The Boolean operators are the NPN class functions from 2 up to $N-1$ inputs, as well as their negated and permuted variants.

For the cost functions, we take area values from LUT macros' bounding box and delay values are the average delay of all timing arcs. If a more sophisticated timing analysis is done, some permutations can generate different delays which can yield different results. This is mitigated at the end of the flow by the SWAP\_PINS capability later presented in Section \ref{subsec:pin_swap_approach}.

The FC-OPT-LUT algorithm takes the number of LUT inputs $N$ and the cost function $C$, which accounts for area and delay. The result is $ALL\_IMP$, a map of functions and LUT implementations. Lines 1-4 initialize the variables $B$, which is the bucket list containing all functions already implemented, and $MAX\_COST$, which will provide the single LUT-$N$ cost. Also, $B$ is initialized with constants and single variables through the method \mbox{CREATE\_INITIAL\_FUNCTIONS}. In line 6, the association of tuples $AT$ is computed, which consists of tuples containing the indices of the buckets used to combine the functions, from index $0$ to $i-1$, where the cost needs to be higher than $B[i-1]$, but at the same time, it is the smaller one of all possibilities. As an example, if the candidate $AT$s have cost $14$, $10$, $10$, $12$ and $B[i-1]$ cost is $9$, the candidate $AT$s with cost $10$ will be selected.

The while loop in lines 7-10 checks, at each iteration, if the cost of $AT$ is not bigger than $MAX\_COST$. In such cases, it is better to use the naive solution. If the cost is smaller, the method $ASSOCIATE$ will process the list of $AT$, combining them 2 by 2 or 3 by 3 (depending on the tuple size), using the cost function for breaking ties. The result is added to the bucket list. A new list of $AT$s is computed which will continue or break the loop. Finally, in lines 12-13, if there are remaining functions, they are considered not decomposable (i.e., the cost to decompose them is higher than the naive solution) and the function CREATE\_NAIVE\_IMPS will look for functions that do not have an implementation on $ALL_IMP$ and add a naive one, guaranteeing that all functions of up to $N$ inputs are present on the $ALL\_IMP$ map.


\begin{algorithm}[tb]
\small
\caption{FC-OPT-LUT Algorithm}
\label{alg:fcopt}
  \SetKwInOut{Require}{Require}
  \SetKwInOut{Ensure}{Ensure}
  \SetKwInput{KwInput}{Input}
  \SetKwInput{KwOutput}{Output}
\DontPrintSemicolon
  \KwInput{$N$ $(number$ $of$ $LUT$ $inputs)$, $C$ $(cost$ $function)$}
  \KwOutput{$ALL\_IMP$}

  $ALL\_IMP \gets \phi$, $B \gets \phi$,   $i \gets 1$
  
  $MAX\_COST \gets $LUT\_COST($C$, $N$)
  
  $B.add$ (CREATE\_INITIAL\_FUNCTIONS ($N$))

  $AT \gets $NEXT\_BUCKET($B$, $i$, $C$)
  
  \While{COST($AT$, $C$) $<$ $MAX\_COST$}
  {
    $B.add$ (ASSOCIATE ($AT$, $C$, $ALL\_IMP$))
  
    $i \gets i + 1$

    $AT \gets $NEXT\_BUCKET($B$, $i$, $C$)
  }
  
  \If{SIZE\_OF($IMP\_LUT$) $<$ $2^{2^N}$}
  {
    CREATE\_NAIVE\_IMPS($ALL\_IMP$, $N$)
  }
  
  \Return $ALL\_IMP$
  
\end{algorithm}

\subsubsection{Heuristic LUT FC method} \label{subsec:fc_heur_alg}


The optimal method described in the previous section is only able to generate LUTs of up to 4 inputs. In order to deal with bigger LUTs, a heuristic is required. A LUT decomposition can be thought of as a factorization problem, where there is a direct conversion from a factored form to a LUT tree (i.e., a fanout-free) structure. With some modifications, we can apply the Boolean factoring method presented in \cite{martins2010boolean} to perform decompositions. This approach is called FC-HEUR-LUT. These modifications are necessary to derive LUT decompositions that can possibly have better cost than the naive solution.

FC-HEUR-LUT is presented in Algorithm \ref{alg:fcheur}. The algorithm takes the target function $F$ and the cost function $C$. The result is the LUT implementation $IMP$, which is the LUT circuit containing the decomposed naive solution. Lines 1-3 initialize the variables $ALL\_IMP$, which represents a map storing all known implementations for the functions already decomposed, and $B$ which contains the buckets. The method \mbox{CREATE\_INITIAL\_FUNCTIONS} remains the same as in FC\_OPT\_LUT. Lines 4-6 check if it is a trivial case and returns if so. In Line 7, the method EXTRACT\_ALL\_COFACTORS is executed, where all the cofactors and cubecofactors (excluding constants) are computed from $F$ and stored in the set $ALL\_COF$. Lines 8-10 are a recursive call to the algorithm, which will provide a LUT implementation to all cofactors and cubecofactors. With the cofactors and cubecofactors derived, the combination of cofactors takes place in Line 12, using the same strategies presented to expand the ``allowed functions'' set, as explained in \cite{martins2010boolean}. This expansion guarantees at least 2 factored subfunctions that when associated in the next step, will provide at least one functionally equivalent solution. In Line 13, the ASSOCIATE\_FUNCTION will perform AND/OR/XOR operations also using the rules mentioned, as well as NAND/NOR/XNOR associations using the ``not comparable'' functions. These associations are discarded if they are not the target function $F$. If the association is functionally equivalent to $F$, the cost function $C$ will compare the current solution (which initially is the naive) with the current one, replacing it in the case of a better cost. Finally, Lines 14-15 will collect the resulting implementation $IMP$ and return.

Some techniques applied to greatly speed-up FC-HEUR-LUT include the usage of FC-OPT-LUT results together with FC-HEUR-LUT. The $ALL\_IMP$ map can be used in the beginning of the algorithm in order to quickly return the optimal implementation if the function $F$ has a support of 4 or less inputs, while also improving the decomposition QoR. Another technique that was applied is a limit on the number of associations of ``not comparable'' functions, because those can be a considerable number (i.e., more than 100 thousand). This avoids considerable runtime trying to decompose more complex functions which generally are not cost worth to decompose.

\begin{algorithm}[t]
\small
\caption{FC-HEUR-LUT Algorithm}
\label{alg:fcheur}
  \SetKwInOut{Require}{Require}
  \SetKwInOut{Ensure}{Ensure}
  \SetKwInput{KwInput}{Input}
  \SetKwInput{KwOutput}{Output}
\DontPrintSemicolon
  \KwInput{$F$ $(target$ $function)$, $C$ $(cost$ $function)$}
  \KwOutput{$IMP$}
  
  $ALL\_IMP \gets \phi$, $B \gets \phi$
  
  $B.add$ (CREATE\_INITIAL\_FUNCTIONS ($F$, $ALL\_IMP$))

  $IMP \gets ALL\_IMP(F)$    

  \If{$IMP \neq \phi$}
  {
    \Return{$IMP$}
  }

  $ALL\_COF \gets$ EXTRACT\_ALL\_COFACTORS ($F$)  

  \ForEach{cofactor $COF \in ALL\_COF$}
  {
      $COF\_IMP \gets$ FC\_HEUR\_LUT($COF$, $C$)
      
      $ALL\_IMP \gets$ [$COF$, $COF\_IMP$]
  }
  
  $ALL\_IMP \gets$ [$F$, GET\_NAIVE\_SOLUTION($F$)]
  
  COMBINE\_COFACTORS($ALL\_COF$, $ALL\_IMP$)
  
  ASSOCIATE\_FUNCTIONS ($ALL\_COF$, $ALL\_IMP$, $C$)
 
  $IMP \gets ALL\_IMP(F)$

  \Return{$IMP$}
  
\end{algorithm}

\vspace{-2mm}
\subsection{Pin Swap Approach} \label{subsec:pin_swap_approach}

The LUTs described in Section \ref{subsec:lut_implementation} can be thought of as a MUX tree fed by registers that store an arbitrary truth table. The MUX tree is the main contributor to the LUT delay, especially in LUTs with a high number of inputs. In this sense, the pin order affects considerably the LUT delay. Paths with small logic depth (i.e., inputs connected to a MUX closer to the output) will be considerably faster.

\begin{figure}[tb]
\centering \footnotesize
\includegraphics[width=1.0\linewidth]{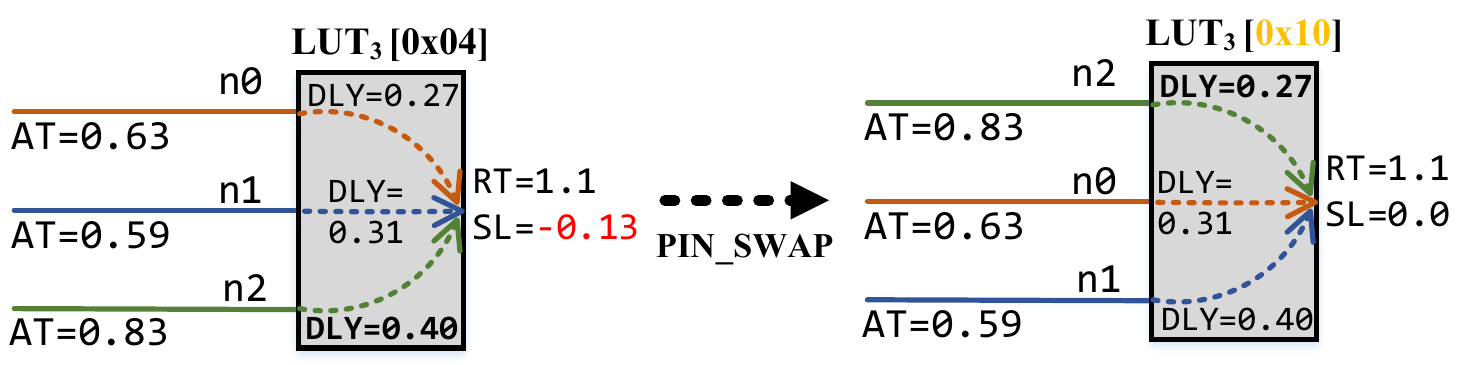}
\caption{Example of a beneficial pin swap}
\centering
\label{fig:pin_swap}
\vspace{-2mm}
\end{figure}

The method SWAP\_PINS takes advantage of the fact that a LUT function can have an arbitrary input pin swap if the truth table is permuted accordingly. So, our method takes a LUT function and its timing information as input and provides the permuted truth table and the new order of input pins/nets. The example presented in Fig.~\ref{fig:pin_swap} shows an effective pin swap that improved the slack of the design. The pin swap algorithm takes the LUT function, the arrival time (AT) of each input net (termed [$n0$, $n1$, $n2$]), the cell arc delay (DLY) associated with each input, and the required time (RT) at the output. In the example, RT=1.1 and the critical arc is $n2$, with a total delay of $1.23$. The algorithm initially tries all the input permutations, trying to minimize WNS. If two or more arcs have negative slack, it also tries to reduce TNS. Once all permutations are tried and there is a new order that improves WNS and/or TNS, the truth table is permuted accordingly to keep the same functionality. The algorithm returns the truth table $0x10$ and the new net order [$n2$, $n0$, $n1$].



\vspace{-2mm}
\section{Experimental Results using TOTe} \label{sec:results}

This section presents the PPA analysis for many selected designs and at different degrees of obfuscation. Without loss of generality, for all experimental results, we have executed FPGA synthesis in Vivado and the target is a Kintex-7 XC7K325T-2FFG900C device which contains 6-input LUTs \cite{ref_8_xilinx}. Following, Cadence Genus is used for the logic synthesis with three flavors of a commercial 65nm standard cell library (LVT/SVT/HVT). However, we emphasize that TOTe is \textbf{agnostic with respect to PDKs, libraries, and tools}.


For our first experiment, we considered a small but pragmatic design which covers all possible FPGA primitives. We selected a schoolbook multiplier (SBM) design as DUO \cite{9417065}. 
We obfuscated an 8-bit SBM by varying $obf_c$ from 55 to 100\% and evaluated the obfuscation versus performance and obfuscation versus area trends. We have synthesized the SBM design targeting a challenging frequency of 540MHz. As calculated by TOTe's timing engine, the $CP$ and $sumCP$ values become 0.490ns and 16088.69ns, respectively. These values correspond to a design obfuscated at 100\%, i.e., all LUTs are reconfigurable. At this stage, these values represent a simplistic timing analysis, realistic timing values will be obtained when the final timing analysis is performed using a commercial physical synthesis tool. However, as we move along with the obfuscation process, $CP$ and $sumCP$ remain consistent in relative terms, which is sufficient to generally determine critical paths to target.

\begin{figure}[tb]
\centering \footnotesize
\includegraphics[width=0.9\linewidth]{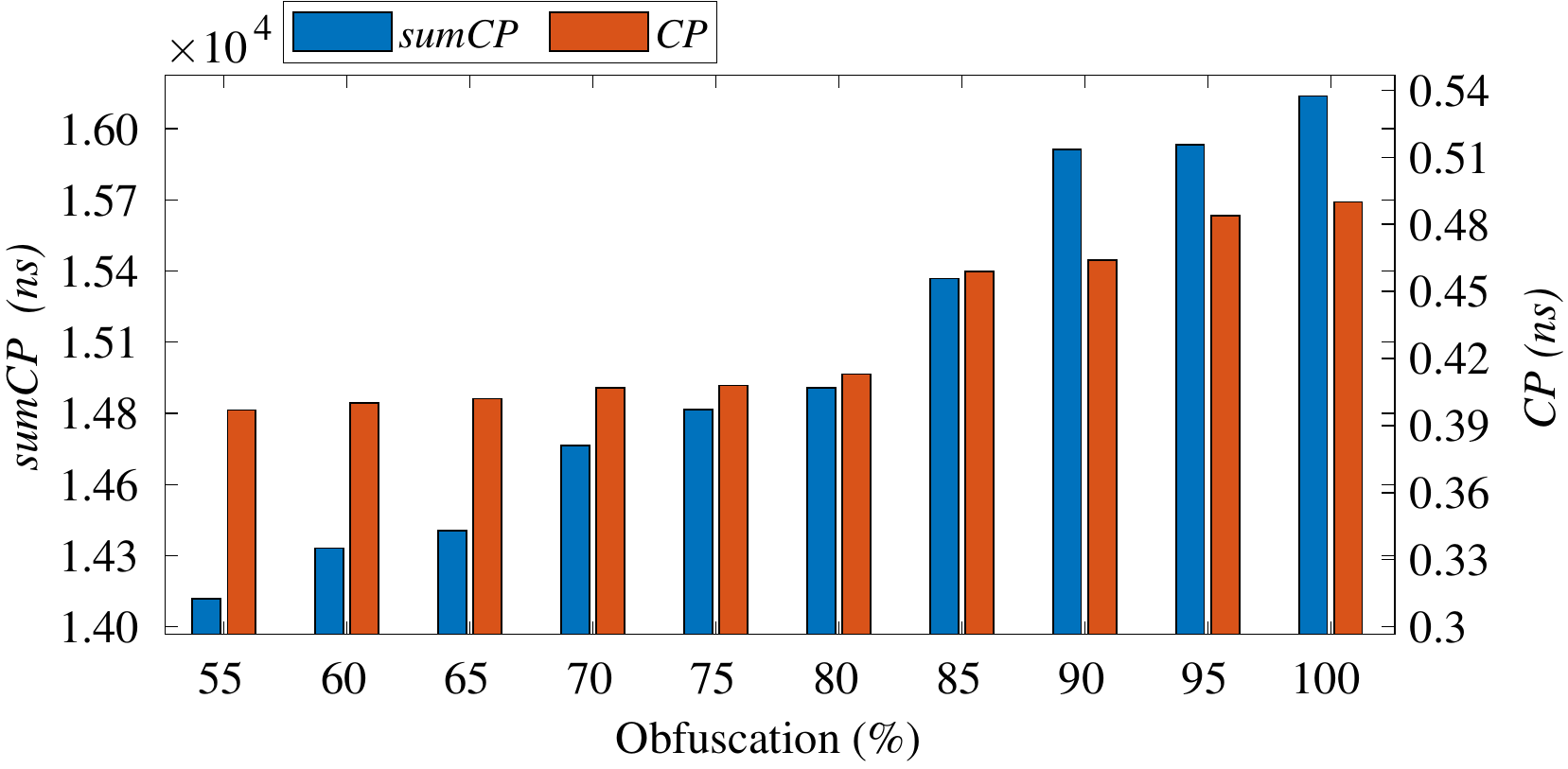}
\caption{TOTe's obfuscation vs. performance for SBM}
\centering
\label{fig:obfuscation_per}
\vspace{-2mm}
\end{figure}
\begin{figure}[tb]
\centering \footnotesize
\includegraphics[width=0.9\linewidth]{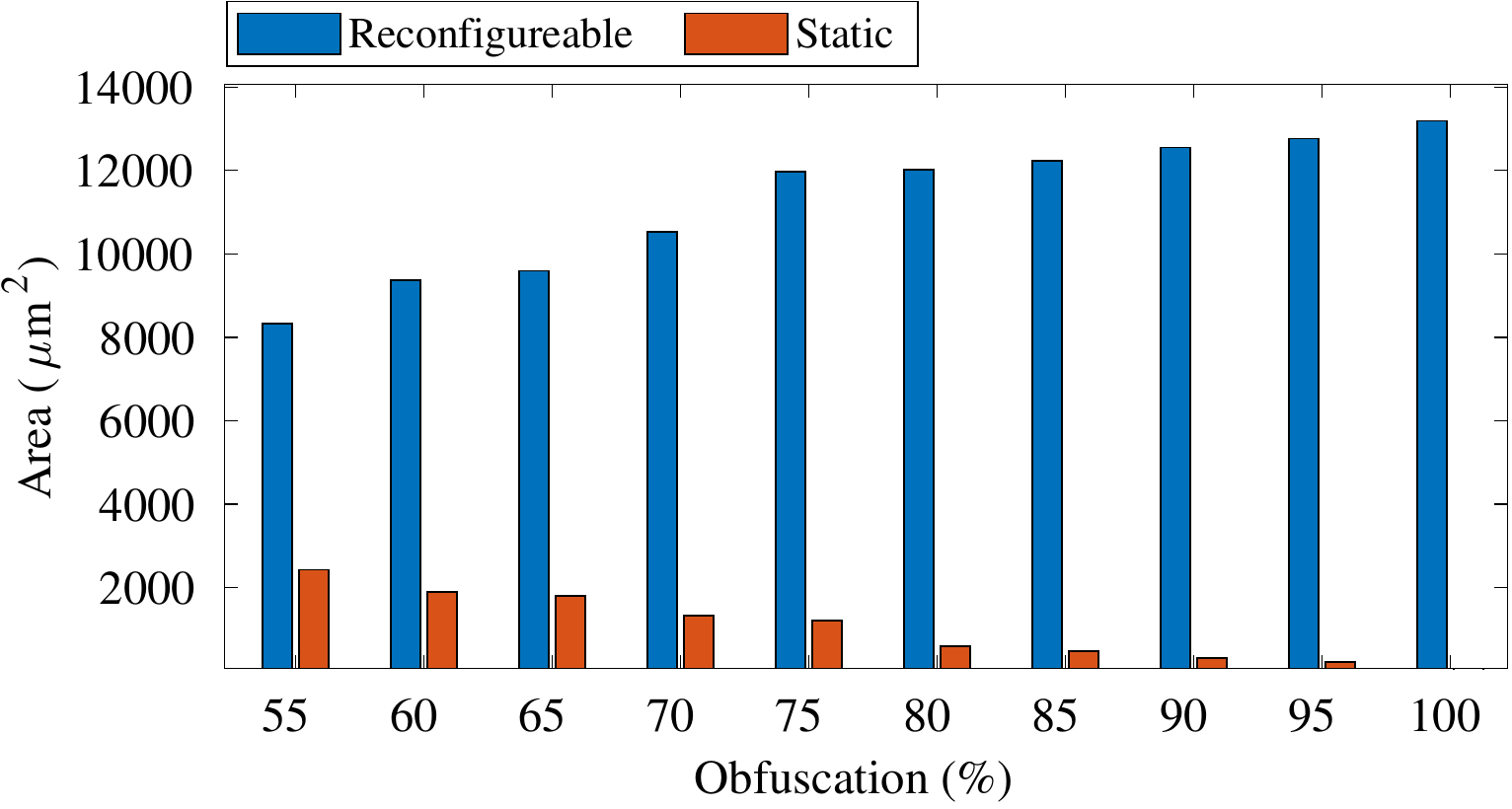}
\caption{TOTe's obfuscation vs. area for SBM}
\centering
\label{fig:obfuscation_area}
\vspace{-2mm}
\end{figure}

After performing the obfuscation for different levels, the timing characteristics for the 8-bit SBM are illustrated in Fig. \ref{fig:obfuscation_per}. The analysis of $CP$ and $sumCP$ shows that performance is decreasing as we increase the level of obfuscation. Conversely, decreasing the level of obfuscation increases the performance of the design. The trend depicted in Fig.~\ref{fig:obfuscation_per} is that $CP$ improves inversely with the obfuscation, but it saturated when the obfuscation is below 80\%. This fact is not true for $sumCP$, the decrease in obfuscation causes continuous improvement as expected. Similarly, the performance versus area profile of the 8-bit SBM is illustrated in Fig.~\ref{fig:obfuscation_area}.

\begin{figure*}[tb]
\centering \footnotesize
\includegraphics[width=0.98\linewidth]{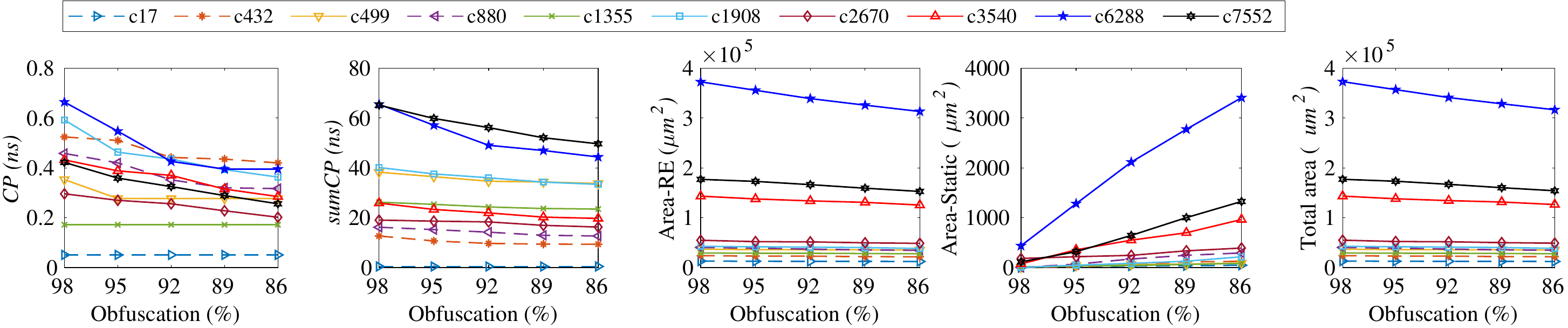}
\caption{Obfuscation results for ISCAS'85 benchmarks}
\centering
\label{fig:obfuscation_results_appendix}
\vspace{-2mm}
\end{figure*}

Next, we wanted to investigate whether the same saturation trend would appear for other designs. We first selected the ISCAS'85 benchmarks and the results are depicted in Fig. \ref{fig:obfuscation_results_appendix}. These relatively outdated benchmarks were selected for the reason that they have a single stage of logic (i.e, they are combinational), so the correlation between $CP$ and $sumCP$ is easy to follow (i.e., the critical path does not change from different reg2reg paths). Even in these simplistic designs, saturation occurs remarkably fast.

\begin{figure*}[tb]
\centering
\includegraphics[width=0.95\linewidth]{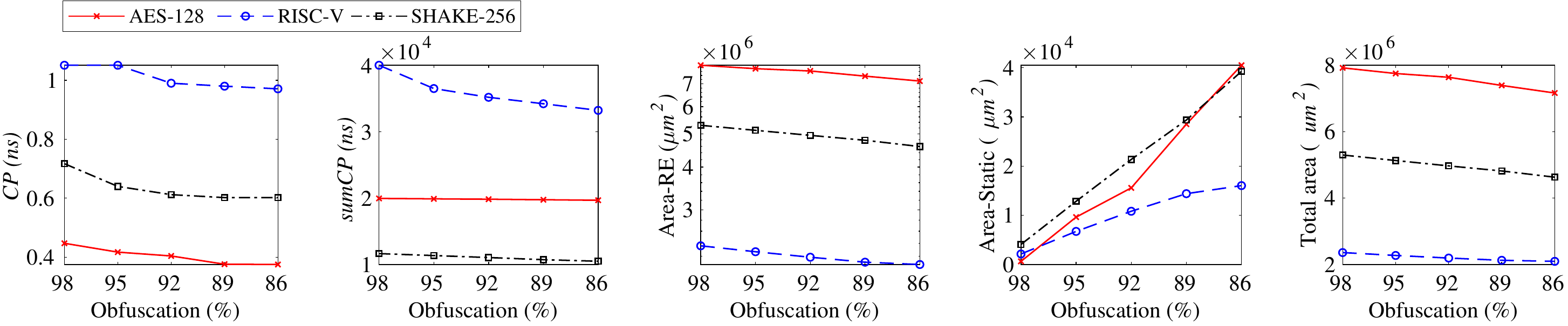}
\caption{Obfuscation results for AES-128, RISC-V and SHAKE-256 using TOTe}
\centering
\label{fig:obfuscation_results}
\vspace{-2mm}
\end{figure*}

Naturally, we have also obfuscated more representative designs. In \cite{eASIC}, detailed results are provided for SBM, SHA-256, and FPU \cite{fpu} designs. For the sake of brevity, we do not repeat those results here. Instead, in Table~\ref{tab:obfuscation_designs}, we provide detailed results for additional circuits, namely IIR, PID, and a GPU. The eight columns of Table~\ref{tab:obfuscation_designs} show design name, obfuscation level, $sumCP$, $CP$, area of the reconfigurable part, area of the static part, number of reconfigurable LUTs, and number of static LUTs. Additional results are also provided in graphical form in Fig.~\ref{fig:obfuscation_results} for AES, RISC-V and SHAKE-256 designs so the trends are easy to visualize.

In summary, the results presented in this section confirm that TOTe is a generic tool for obfuscation and it can obfuscate a design regardless of its complexity. It also becomes clear that the reliance on a LUT-based representation of the circuit, akin to an FPGA, has different implications for delay and area. For area, the trend is clear: the lesser is the obfuscation target, the more compact the circuit becomes. However, for delay, it appears that eASIC brings performance penalties that cannot be overcome by simply reducing the targeted obfuscation level. Therefore, other strategies are needed for achieving better performance. In the next section, we will present a more detailed analysis during physical synthesis. We will also apply the optimization methods described in Section \ref{sec:dec_alg} for improving performance.

\begingroup
\setlength{\tabcolsep}{2.0pt} 
\begin{table} [tb]
\footnotesize \centering
\caption{Detailed results for selected designs using TOTe}
\label{tab:obfuscation_designs}
\begin{tabular}{|p{1.18cm}|p{0.515cm}|p{1.2cm}|p{0.7cm}|p{1.35cm}|p{1.20cm}|p{0.75cm}|p{0.70cm}|}
\hline 
\textbf{Design} & \textbf{Obf. (\%)} & \textbf{\textit{sumCP} \, ($ns$)} & \textbf{\textit{CP} ($ns$)} & \textbf{Area-RE ($\mu m^2$)} & \textbf{Area-ST ($\mu m^2$)} & \textbf{LUT\, (RE)} & \textbf{LUT \, (ST)} \\ \hline \hline
\multirow{5}{*}{IIR \cite{iir}} & 98 & 1574.48 & 0.591 & 55031.04 & 257.4 & 584 & 11  \\ \hhline{~-------}
& 95 & 1553.32 & 0.526 & 54104.40 & 720.72 & 566 & 29 \\ \hhline{~-------}
& 92 & 1534.39 & 0.526 & 53177.76 & 1184.04 & 548 & 47 \\ \hhline{~-------}
& 89 & 1501.29 & 0.526 & 52251.36 & 1647.36 & 530 & 65 \\ \hhline{~-------}
& 86 & 1489.93 & 0.526 & 51324.48 & 2110.68 & 512 & 83 \\ \hline \hline
\multirow{5}{*}{PID \cite{pid}} & 98 & 2547.58 & 0.756 & 445590.00 & 2816.82 & 896 & 18  \\ \hhline{~-------}
& 95 & 2466.25 & 0.642 & 432340.92 & 9441.36 & 869 & 45 \\ \hhline{~-------}
& 92 & 2391.96 & 0.592 & 421365.95 & 14928.84 & 841 & 73 \\ \hhline{~-------}
& 89 & 2348.61 & 0.568 & 407273.76 & 21974.94 & 814 & 100 \\ \hhline{~-------}
& 86 & 2322.46 & 0.543 & 392345.64 & 29439.00 & 787 & 127 \\ \hline \hline
\multirow{5}{*}{\parbox{1.45cm}{GPU (OR1200-HP) \cite{open_risc}}} & 98 & 237699.30 & 0.933 & 21317740.2 & 124971.48 & 40739 & 831  \\ \hhline{~-------}
& 95 & 215696.68 & 0.871 & 21009821.4 & 278931.10 & 40102 & 2078 \\ \hhline{~-------}
& 92 & 185520.65 & 0.750 & 20015822.2 & 495521.11 & 39492 & 3352 \\ \hhline{~-------}
& 89 & 154560.56 & 0.650 & 19552256.6 & 781521.30 & 38243 & 4521 \\ \hhline{~-------}
& 86 & 135802.32 & 0.625 & 18552023.3 & 1011230.2 & 36125 & 5806 \\ \hline

\end{tabular}
\end{table}
\endgroup


\vspace{-2mm}
\section{Physical Synthesis for eASIC} \label{sec:physical_implementation}

This section contains the physical implementation results for an obfuscated SHA-256 \cite{sha256} design. Cadence Innovus is utilized for physical synthesis, together with a commercial 65nm PDK. We have selected SHA-256 as it is popular and widely used in cryptography. The variants of the design with different obfuscation levels are implemented with the aid of the LUTs defined in Section~\ref{subsec:lut_implementation}. The results obtained after implementation are focused on performance vs. area trade-offs for the 80-100\% obfuscation range, thus avoiding the saturation trend highlighted in Section \ref{sec:results}.

Initially, we synthesized and implemented SHA-256 on FPGA, the target device being a Kintex-7. The FPGA implementation achieves a frequency of only 77 MHz (for reference, the Kintex-7 family is produced on a 28nm CMOS technology). To start the analysis, we select 100\% obfuscation as a baseline design because it is fully reconfigurable and somewhat analogous to an FPGA design. The implementation results for 80\%, 85\%, 90\%, and 100\% obfuscation are given in Table \ref{tab:sha_impl}. The timing results are obtained after physical synthesis and are for the worst process corner (SS), VDD=$0.9*VDD_{nominal}$, and a temperature of $125^{\circ}$C.


From the results, it is clear that the level of obfuscation does not affect the utilization density of the design (i.e., ratio of placement sites that are occupied vs. empty). For all designs, we achieved around 80\% utilization density, which is very high considering the large number of macros. In other words, our macros do not compromise global routing resources. It is noteworthy that the performance of TOTe-generated designs is increasing as we decrease the level of obfuscation; our baseline eASIC design is running at 223 MHz (as shown in Freq. column of Table \ref{tab:sha_impl}) and it increases as $obf_c$ decreases. This behavior matches the goal we set from the start: to establish a trade-off between performance (ASIC) and security (FPGA).


\begingroup
\setlength{\tabcolsep}{2.0pt} 
\renewcommand{\arraystretch}{1.1} 
\begin{table*} [!htb]
\footnotesize \centering
\caption{Results for the implementation of SHA-256 for different obfuscation levels}
\label{tab:sha_impl}
\begin{tabular}{|p{1.1cm}|p{1.0cm}|p{1.0cm}|p{1.5cm}|p{1cm}|p{1cm}|p{1.1cm}|p{0.8cm}|p{1cm}|p{1cm}|p{0.7cm}|p{1.5cm}|p{1.4cm}|}
\specialrule{0.15em}{0em}{0em} 

\textbf{CAD flow} & \textbf{Obf.} & \textbf{Density} & \textbf{Area ($\mu m^2$)} & \textbf{Freq. ($MHz$)} & \textbf{Leakage ($mW$)} & \textbf{Dynamic Power ($mW$)} & \textbf{\# LUT} & \textbf{\# Buffer}  & \textbf{\# Comb.} & \textbf{\# Inv.} & \textbf{\# Sequential} & \textbf{Total  Wirelength ($\mu m$)} \\ \specialrule{0.15em}{0em}{0em} 
FPGA & 100\% & -- & -- & 77 & 2.4 & 191 & 2238 & -- & -- & -- & 1830* & --  \\ \specialrule{0.15em}{0em}{0em} 

TOTe & 100\% & 81\% & 1751500 & 223 & 14.85 & 505.05 & 2238 & 5846 & 93470 & 6175 & 105128 & 9247654\\ \hline
TOTe & 90\%  & 77\% & 1638500 & 234 & 12.23 & 438.47 & 2015 & 4626 & 84107 & 5017 & 94876 & 7505590 \\ \hline
TOTe & 85\%  & 80\% & 1507000 & 241 & 12.10 & 430.98 & 1904 & 4846 & 80304 & 5585 & 90420 & 7207023  \\ \hline
TOTe & 80\%  & 80\% & 1409700 & 248 & 11.05 & 386.89 & 1792 & 4406  & 75083 & 4564 & 83790 & 6724434  \\ \specialrule{0.15em}{0em}{0em} 
ASIC & NONE & 92\% & 34208 & 248 & 0.18 & 9.37 & -- & 167 & 3244 & 190 & 1812 & 158003\\ \hline
ASIC & NONE  & 91\% & 40804 & 550 & 0.299 & 23.86 & -- & 722 & 3244 & 956 & 1812 & 181441 \\  \specialrule{0.15em}{0em}{0em} 
\end{tabular}
\begin{tablenotes}
      \centering
      \item  \footnotesize * Vivado performs flip-flop cloning for solving high fanout buffering. Thus, there is an increase in the number of registers w.r.t. ASIC.  
    \end{tablenotes}
\vspace{-2mm}
\end{table*}
\endgroup

\begin{figure*}[tb] \centering \footnotesize
\includegraphics[width=\linewidth]{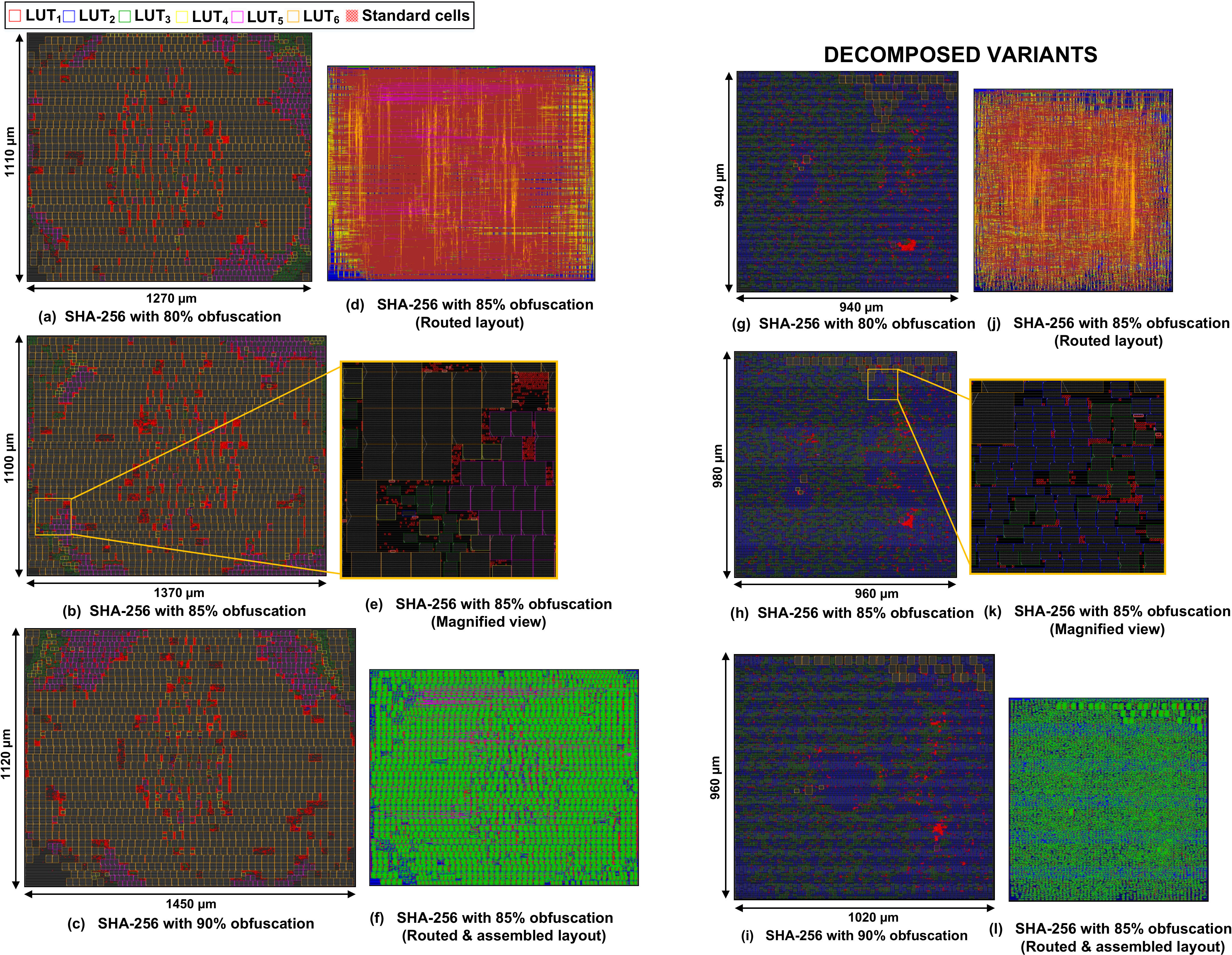}
\caption{Implementation results for SHA-256 with different obfuscation levels}
\centering
\label{fig:obfuscation_impl}
\vspace{-2mm}
\end{figure*}


The area of the design is proportional to the obfuscation level, which means that increasing the security of design comes with an area penalty. As we only exploit LUT primitives for promoting obfuscation, the number of LUTs increases with the obfuscation level. In the same manner, leakage and dynamic power figures are proportional to security as reconfigurable logic is less efficient than static. This is mainly because of the use of flip-flops to store the LUT truth tables. The last five columns of Table \ref{tab:sha_impl} show the resource requirements for eASIC (number of buffers, combinational cells, inverters, sequential cells, and the total wirelength). 

In Fig. \ref{fig:obfuscation_impl}, we show many different views of the SHA-256 layouts under different obfuscation targets. The considered metal stack has 7 metals assigned to signal routing. Panels (a)-(c) of Fig. \ref{fig:obfuscation_impl} illustrate the layouts for 80\%, 85\% and 90\% obfuscation levels. The dimensions of the layouts are indicated on the bottom and left sides of each panel. All the six variants of LUTs are highlighted with different colors and the static part of eASIC is highlighted in red -- notice that, as expected, the design remains primarily a sea of LUTs. The majority of those LUTs are LUT\textsubscript{6}, thus the layouts appear to be dominated by yellow boxes.

Panel (d) of the same figure demonstrates the final post route layout of eASIC. Notice how the post route design contains mostly vertical orange lines which correspond to M6. Panel (e) of Fig. \ref{fig:obfuscation_impl} shows the magnified view of the placement in an eASIC design. The mixed structure of LUT macros and standard cells clearly depicts the placement pattern and spacing between the macros is usually filled with standard cells. Notice how the LUT macros align with the standard cell rows, allowing for the entire design to have a uniform power rail and power stripe configuration. In panel (f) of the same figure, we illustrate the same design but filter out some routing layers (only M2, M3 and M4 are shown). As depicted in Fig. \ref{fig:lut_imp}, the implemented LUTs utilize the aforementioned metal layers, therefore the assembled view of panel (f) represents how visually regular the eASIC structure is.

In the results shown in panels (g)-(l) of Fig. \ref{fig:obfuscation_impl}, we have utilized the same design and conditions but applied LUT decomposition for improving performance. Notice that the layouts are drawn to scale to highlight the area reduction brought by decomposition. In particular, when comparing panels (f) and (l), we note that regularity is still present even after decomposition, as expected.

The detailed results for these designs are listed in Table~\ref{tab:sha_impl_decomposed}. With decomposition, the baseline frequency increased significantly, from 223 to 307MHz. As in the non-decomposed version, the performance increases inversely with the obfuscation level. On top of that, the area was reduced by more than half, along with the power consumption. However, due to a large number of small LUTs (mostly LUT\textsubscript{2}s), placing and routing becomes slightly more challenging. For this reason, the maximum utilization density across the optimized designs is approximately 65\%. Nevertheless, decomposition is very beneficial: the gain in PPA when compared with the non-optimized versions is significant. We argue that since decomposition has a negligible impact on the runtime of the physical synthesis flow, it should always be applied. For instance, the runtime to apply the decompositions in the SHA-256 circuit with 100\% obfuscation containing 2238 LUTs was 11 minutes in an Intel Core i7-6700K. The decomposition achieved improvements of 50\% in the LUT area and 32\% in the total LUT delay.   


While the LUT decomposition brings significant performance improvement, we seek to achieve performance levels that are as close to the ASIC implementation as possible. For that reason, after the decomposition, we also applied pin swapping technique. As we noted earlier, this is possible because the same logic function can be generated with different input orders and different masking bits (truth table). Therefore, we can search for LUTs that appear on the critical path(s) and swap their pins to reduce the total delay. 

\begin{figure}[tb]
\centering \footnotesize
\includegraphics[width=\linewidth]{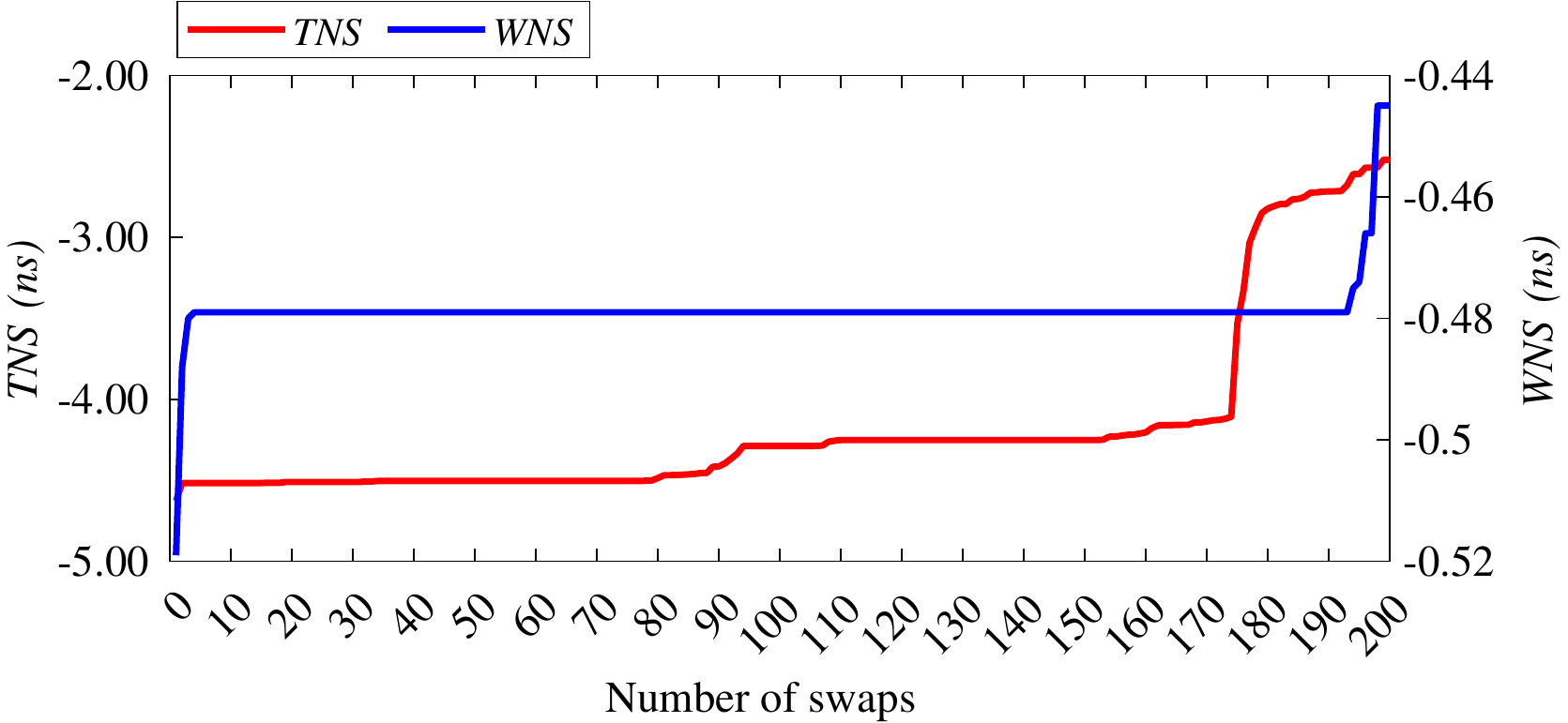}
\caption{Change in the TNS/WNS with respect to the swap of LUT pins}
\centering
\label{fig:obfuscation_impl_decomposed_swap}
\vspace{-2mm}
\end{figure}

For illustrating the capability of pin swapping, we first create an artificial scenario where we increase the target frequency of the design until several paths violate setup timing. The frequency increase determines the number of violating paths that indirectly determines the number of LUTs that will be considered for pin swap purposes\footnote{Here we establish a runtime vs. QoR trade-off. The more aggressive the frequency target is, the more LUTs are considered for pin swap.}. All LUTs from violating paths are chosen as candidates and saved in a list. Then, iteratively, starting with the worst violating path, the pins of the LUTs are swapped until the critical path is improved (i.e., WNS). Number of swaps performed versus the TNS and WNS is illustrated in Fig.~\ref{fig:obfuscation_impl_decomposed_swap}. From this figure, the first few swaps improved the WNS while the TNS remain the same. The continuing swapping start to improve the TNS without any change in the WNS. If the TNS is improving, that means there is a chance to reach a better WNS. Thus, we performed the swapping until the next jump in the WNS. After attempting 200 swaps, WNS improved by approximately 80ps and TNS by 2ns, thus making the design 11MHz faster. 



\begingroup
\setlength{\tabcolsep}{2.0pt} 
\renewcommand{\arraystretch}{1.1} 
\begin{table*} [!htb]
\footnotesize \centering
\caption{Results for the implementation of SHA-256 for different obfuscation levels with decomposed LUTs}
\label{tab:sha_impl_decomposed}
\begin{tabular}{|p{1.6cm}|p{1.0cm}|p{1.0cm}|p{1.5cm}|p{1cm}|p{1cm}|p{1.1cm}|p{1cm}|p{0.8cm}|p{1cm}|p{1cm}|p{0.7cm}|p{1.4cm}|}
\hline 
\textbf{CAD flow} & \textbf{Obf.} & \textbf{Density} & \textbf{Area ($\mu m^2$)} & \textbf{Freq. (MHz)} & \textbf{Leakage ($mW$)} & \textbf{Dynamic Power ($mW$)} & \textbf{\#LUT} & \textbf{\#Buf.}  & \textbf{\#Comb.} & \textbf{\# Inv.} & \textbf{\#Seq.} & \textbf{Total Wirelength ($\mu m$)} \\ \hline
TOTe & 100\% & 61\% & 1155000 & 307 & 8.00 & 301.49 & 10182 & 3583 & 29352 & 15261 & 53868 & 3391742  \\ \hline
TOTe & 90\%  & 65\% & 979200 & 312 & 7.55 & 273.54 & 9127 & 1797 & 27115 & 13538 & 49016 & 3242970  \\ \hline
TOTe & 85\%  & 67\% & 940800 & 322 & 7.03 & 256.36 & 8676 & 1882 & 26011 & 13136 & 46796 & 2982627  \\ \hline
TOTe & 80\%  & 64\% & 883600 & 357  & 6.44 & 278.37 & 8124 & 1726  & 24614 & 12340 & 43830 & 2889253  \\ \hline
TOTe (Swap) & 80\%  & 64\% & 883600 & 368  & 5.93 & 283.35 & 8124 & 1726  & 24614 & 12340 & 43830 & 2889760  \\ \hline
\end{tabular}
\vspace{-2mm}
\end{table*}
\endgroup


\vspace{-2mm}
\section{Security Analysis} \label{sec:security_analysis}

On the previous section, we have detailed the design trade-offs associated with an eASIC solution. This section explores the security concerns with eASIC, which are associated with attack vectors stemming from the portion of the design that is exposed (from an adversary's point of view).


\vspace{-2mm}

\subsection{Threat Model} \label{subsec:threat_model}
In our considered threat model, the primary adversary is the \emph{untrusted foundry}. We make no distinction whether the adversary is institutional or a rogue employee. Assuming the security of an eASIC design is a function of its static logic (fully exposed) and reconfigurable logic (protected by a bitstream that serves as a key), we make the following assumptions:

\begin{itemize}
    \item The main adversary goal is to reverse engineer the design in order to pirate its IPs, overproduce the IC, or even to insert sophisticated hardware trojans. For this goal, the adversary \textbf{must} recreate the bistream.
    
    \item The adversary goal might also be to identify the circuit intent, even in the presence of obfuscation. For this goal, the adversary \textbf{does not need} to recreate the bistream.
        
    \item The adversary has access to the GDSII file of the eASIC design sent for fabrication. The adversary is skilled in IC design and has the knowledge and tools required for understanding this layout representation.
    
    \item The adversary can recognize the standard cells, therefore the gate-level netlist of the obfuscated circuit can be easily recovered \cite{regds}.
    
    \item The attacker can identify reconfiguration pins \cite{eval_logic, yasin_logic_locking}, thus being able to effortlessly enumerate all LUTs and their programming order.
    
    \item The adversary can group the standard cells present in the static logic and convert them back into a LUT representation\footnote{This is a very generous concession since the static logic is repeatedly optimized during logic and physical synthesis. Nevertheless, we err on the side of caution and assume the adversary can achieve a perfect reconstruction of LUTs, which by itself is a reverse engineering problem.}.
    


    
    
    
\end{itemize}

We have proposed two different attacks to evaluate the security hardness of eASIC: one based on the \emph{structure} of design and another based on the \emph{composition} of known different circuits. We assert that an adversary can learn and extract information by exploiting the static portion of the design, including the frequency of specific masking patterns. This capability would allow for an adversary to shrink the search space for the key that unlocks the design. 

Similarly, the notion of masking pattern frequency can be utilized as a template to compare different designs. In other words, the composition of the LUTs in a design would allow for a template-based attack. Moreover, we have also evaluated the security hardness of eASIC for conventional oracle-guided and oracle-less attacks borrowed from logic locking attacks. All the experiments reported in this section were run on a server equipped with 32 processors (Intel(R) Xeon(R) Platinum 8356H CPU @ 3.90GHz) with 1.48TB of RAM.



\vspace{-2mm}

\subsection{Structural Analysis Attack} \label{subsec:structural_analysis}
\noindent \textbf{Goal}: by statistical analysis means, decrease the key search space before attempting to recover the bitstream.

We recall again that TOTe's obfuscation engine utilizes six variants of LUTs (LUT\textsubscript{1}, LUT\textsubscript{2}, ..., LUT\textsubscript{6}). However, the majority of the LUTs are LUT\textsubscript{6} due to the packing algorithm executed during FPGA implementation. Therefore, in our security analysis, let us assume without loss of generality, that any FPGA-synthesized circuit contains only LUT\textsubscript{6} instances. 

For a LUT\textsubscript{6}, the possible number of keys is $2^{64}$. But this number is only realistic if the FPGA synthesis tool is genuinely able to exercise the entire key search space. This does not appear to be true: We have synthesized a considerable number of representative designs ($>$30) and extracted all unique LUT\textsubscript{6} masking patterns from the corresponding netlists. We term these values $m_i$. We considered designs of varied size, complexity, and functionality until the combined number of unique masking patterns forms a set of $M = \sum m_i = 3376$ elements that appears to settle. This result alone, albeit being empirical, reduces the global search space from  $2^{64}$ to $3376=2^{11.72}$.

\begin{figure*}[tb]
\centering \footnotesize
\includegraphics[width=0.89\linewidth]{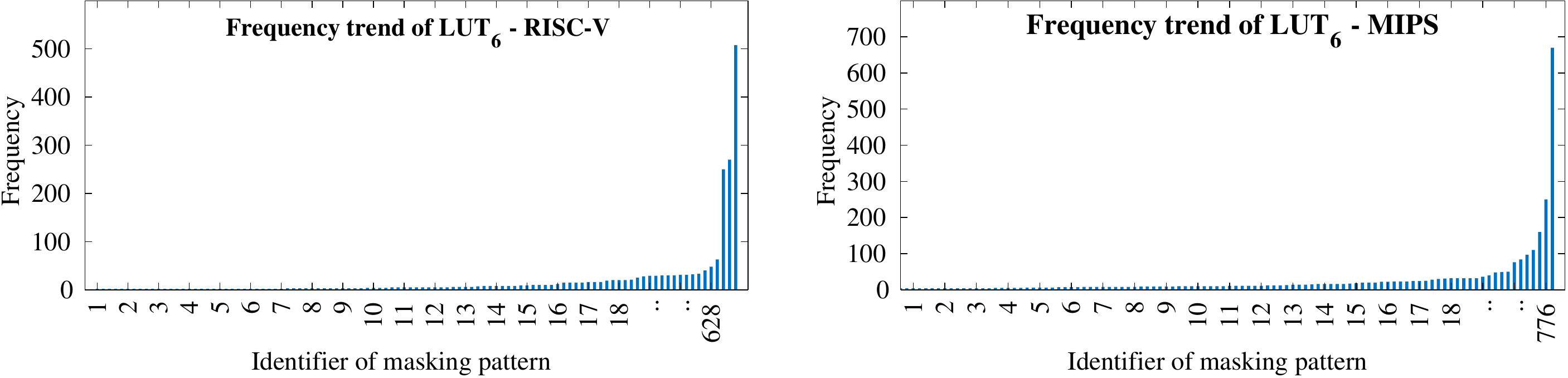}
\caption{Frequency of masking patterns for RISC-V and MIPS}
\centering
\label{fig:obfuscation_bar}
\vspace{-2mm}
\end{figure*}

With this information at hand, we hypothesize that an attacker can exploit the frequency at which LUTs appear in a netlist in order to mount attacks, thus the name structural analysis attack. In other words, the adversary is interested in finding the value of $m_i$ for a given circuit $C_i$. However, the adversary only has partial knowledge about the design. The question then becomes whether the adversary can estimate $m_i$ by performing statistical analysis on a portion of $C_i$. To this end, we targeted two processor designs in our statistical analysis: MIPS and RISC-V. For each circuit, we utilize tuples of $\langle pattern, frequency \rangle$ for tracking how often masking patterns repeat. The pattern is a 64-bit hexadecimal number. The tuples are referenced by integer identifiers and ordered by frequency as shown on the bar charts in Fig. \ref{fig:obfuscation_bar}. Notice that the MIPS netlist has 776 unique LUTs and that there are very few outliers that occur more than 50 times. Similarly, for RISC-V, there are 628 unique LUTs and only 3 occur more than 100 times. 

\begin{figure*}[ht]
\centering \footnotesize
\includegraphics[width=0.89\linewidth]{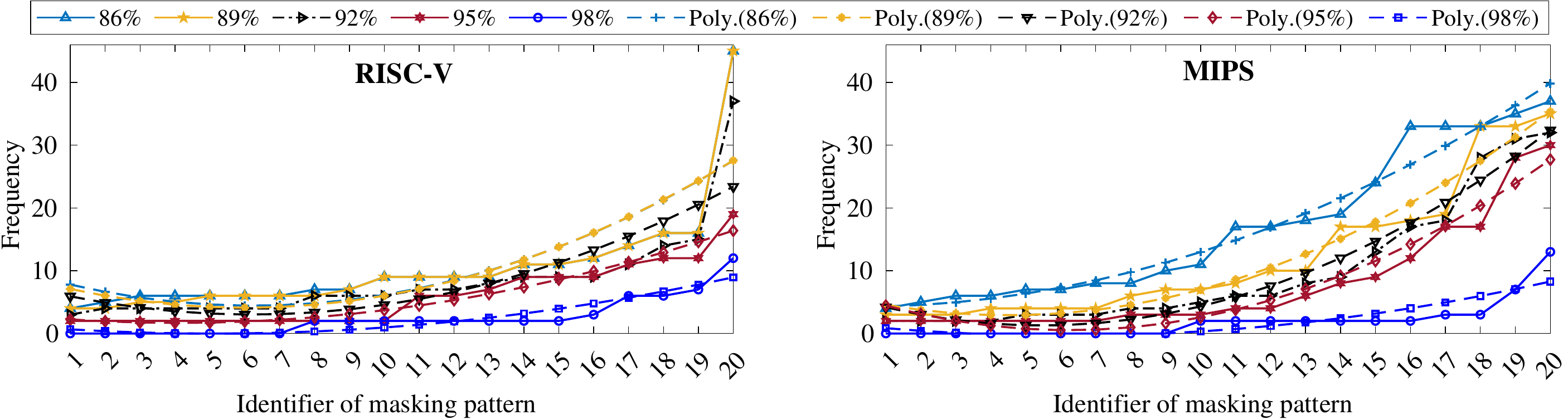}
\caption{The structural analysis of RISC-V and MIPS}
\centering
\label{fig:obfuscation_distr}
\vspace{-2mm}
\end{figure*}

We investigate this by analysing the behaviour of the frequency of masking patterns as depicted in Fig.~\ref{fig:obfuscation_distr}. For this, we utilized netlists generated by TOTe at 98\%, 95\%, 92\%, 89\%, and 86\% obfuscation levels. Therefore, for this experiment, we assume the attacker only has visibility over 2\%, 5\%, 8\%, 11\%, and 14\% of the LUTs, respectively. The adversary then attempts to predict the distribution of actual masking patterns in the design from his/her observation of the small percentage of LUTs that are exposed in the static portion of eASIC. In Fig. \ref{fig:obfuscation_distr}, the adversary's guessing attempt is performed with the aid of polynomial trendlines. For MIPS and RISC-V, it appears that the adversary can estimate to some degree what masking patterns are the outliers. The actual number of unique patterns, $m_i$, is not trivial to determine from extrapolation since many patterns appear a single time or very few times (see Fig. \ref{fig:obfuscation_bar}). We clarify that several circuits studied in this paper (e.g., PID, IIR, GPU) have a similar profile, where only a handful of high-frequency LUTs appear. Therefore, it remains to be studied if the knowledge gathered from this attack can be useful for some form of hill climbing attack (or even a biased version of SAT). 






\vspace{-2mm}
\subsection{Composition Analysis Attack} \label{subsec:composition_analysis}
\noindent \textbf{Goal}: identify the circuit by correlation to known circuits.

This attack also exploits the frequency of the LUTs, but here we correlate several designs against each other based on their composition, thus the name. We suppose that the attack is already successful if the adversary is able to identify the circuit (i.e., breaking the key is not necessary).

\begin{figure*}[tb]
\centering 
\includegraphics[width=0.95\linewidth]{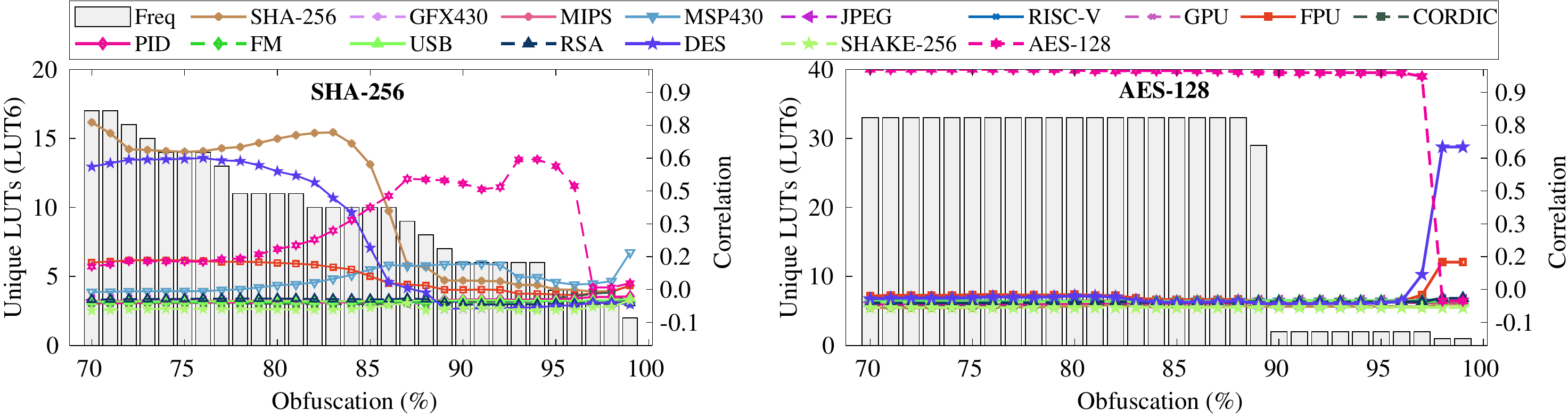}
\caption{The correlation of SHA-256 and AES-128 versus numerous other designs}
\centering
\label{fig:corr_sha}
\vspace{-2mm}
\end{figure*}

In the experiment depicted in Fig. \ref{fig:corr_sha}, we carried out correlation analysis for two different crypto cores: SHA-256 and AES-128. The goal of this analysis is to examine the leaked information from the static part against a database\footnote{We assume the adversary can obtain samples of open source cores from repositories and execute FPGA synthesis on them to create a database.} of circuits that are known to the attacker. We perform obfuscation of SHA-256 and AES-128 in the 70-100\% range and then correlate their static portions with the designs in the database. The x-axis of Fig. \ref{fig:corr_sha} is the obfuscation level, the y-axis is the number of unique LUTs (left) and Pearson correlation (right).


The correlation results reveal very interesting trends. For SHA-256, three regions of interest can be defined based on the degree of obfuscation: 97-100\% (no correlation), 86-96\% (strong correlation to another circuit), and 70-85\% (correlation to itself). A similar analysis has been performed for AES-128 as shown in the right side of Fig. \ref{fig:corr_sha}. The correlation between obfuscated AES-128 versus AES-128 is almost one for obfuscation $<$ 97\%. Conversely, the correlation for obfuscated AES-128 versus other designs is almost zero obfuscation in the same range ($<$ 97\%). 

Assuming that the adversary's objective is exclusively to recognize the circuit's intent (``what is this circuit?''), eASIC could prove as vulnerable as an ASIC design. This is the case for the AES-128 circuit, while the SHA-256 case reveals a contrasting trend: there are obfuscation ranges that can be targeted on purpose to confuse an adversary. For SHA-256, this range appears to be 86-96\%. 

Finally, for an adversary that is interested in obtaining the bitstream, we hypothesize that the correlation analysis herein depicted might be useful to shrink the key search space further. In practice, if the attacker could know for a fact that the obfuscated circuits are indeed AES, or SHA-256, or $C_i$, his key guessing would be based on the $m_i$ of the circuit with the highest correlation. Referring again to the example from the previous subsection, the search space would shrink further; from 3376 to 776 for MIPS and from 3376 to 628 for RISC-V. From this point onward, in order to obtain the key to unlock a design, an adversary would still have to resort to other attacks that are not specific to eASIC. We discuss such attacks in the subsections that follow.

\vspace{-2mm}
\subsection{Oracle-Guided attacks} \label{sbsec:sat_attacks}
\noindent \textbf{Goal}: to retrieve a key or a key guess.

As compared to conventional logic locking \cite{intro_epic}, the LUTs introduced in eASIC are the elements that serve as key gates. A LUT\textsubscript{6}, in theory, introduces 64 bits of key, akin to 64 XOR/XNOR gates in conventional logic locking. The very first circuit we introduced in Section \ref{sec:results}, the SBM, has 25 LUTs (out of which 11 are LUT\textsubscript{6}) when its obfuscation rate is 86\%. In turn,  the key search space would be $2^{11\times64}$ for LUT\textsubscript{6} alone. 

Such a large search space would discourage an adversary from performing SAT attacks on eASIC. However, enumerating the key search space is a very simplistic/naive approach. One has to perform actual attacks in order to evaluate the security of the designs, especially well-known satisfiability-based attacks. We have therefore considered three different SAT attacks to evaluate the security hardness of eASIC: Conventional SAT \cite{SAT}, AppSAT \cite{AppSAT}, and ATPG-based SAT \cite{ATPG}. All three considered attacks operate on .bench files and only take combinational circuits as input. For this reason, we built a script that converts eASIC designs by replacing flip-flops that store key bits for primary inputs. 

\begin{figure}[tb]
\centering 
\includegraphics[width=\linewidth]{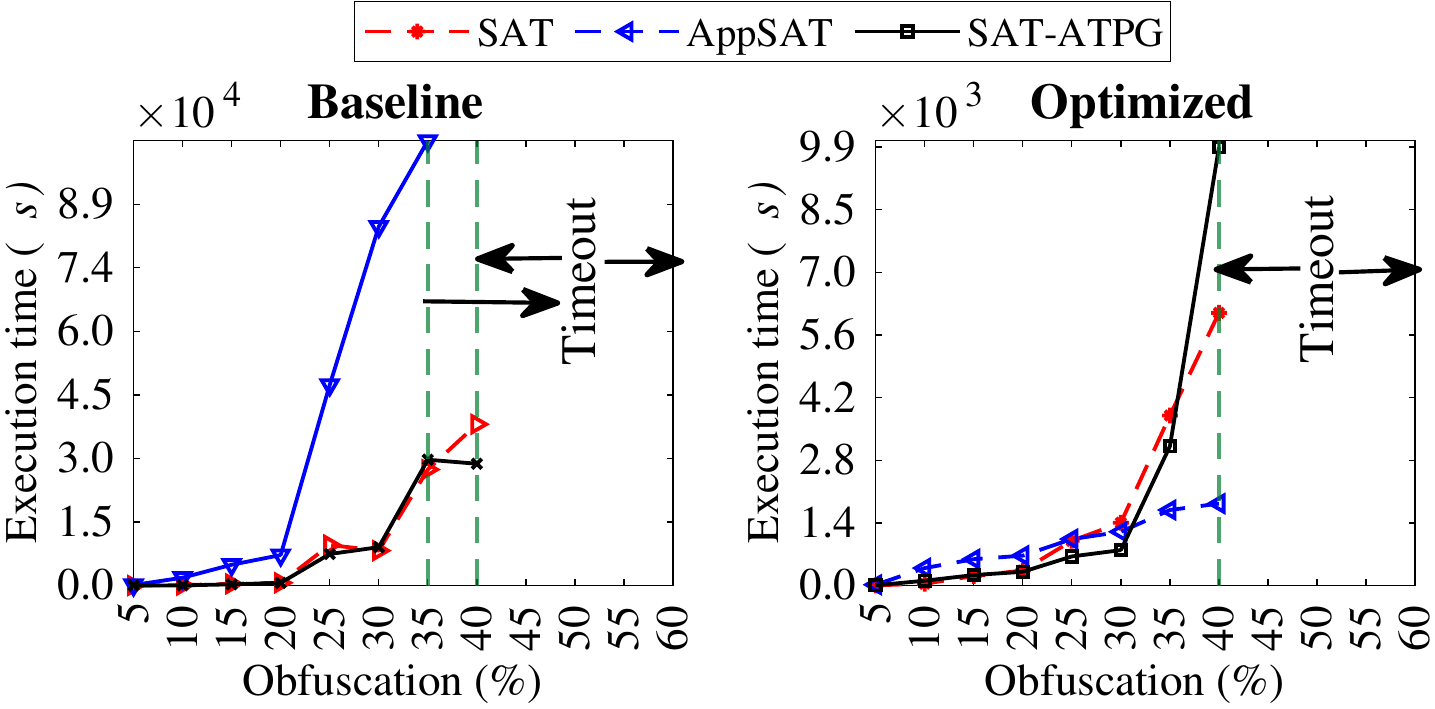}
\caption{Comparison between the baseline and optimized design for the execution time of SAT attacks.}
\centering
\label{fig:sat_time}
\vspace{-2mm}
\end{figure}

We have selected \emph{c7552}, a large but representative design from the ISCAS'85 benchmarks to evaluate the security hardness of eASIC against SAT-based attacks. Importantly, we present the results for two different variants of eASIC, optimized and non-optimized. TOTe automatically optimizes designs by decomposing LUTs into smaller LUTs. While power, area, and performance is improve, the decomposition reduces the size of the key for unlocking the design. We must therefore verify that the reduction in key size does not make the eASIC designs vulnerable to existing attacks. Fig. \ref{fig:sat_time} illustrates the execution time for different SAT attacks and different obfuscation rates. The x-axis of the figure shows the obfuscation level and the y-axis shows the corresponding execution time. As expected, the execution time increases as we increase the obfuscation level. The region to the left of the red line shows successful SAT attacks. However, the region on the right corresponds to unsuccesful attacks, where a timeout of 48h was achieved before the solver returned an answer. In principle, this is an encouraging result since even a very small and combinational-only circuit like \emph{c7552} leads to timeouts at relatively low obfuscation rates. For the optimized version, we can see that there is a reduction in the time that it takes for the successful attacks to complete. However, no attacks is sucesful beyond 40\% obfuscation.




\begin{figure}[tb]
\centering 
\includegraphics[width=\linewidth]{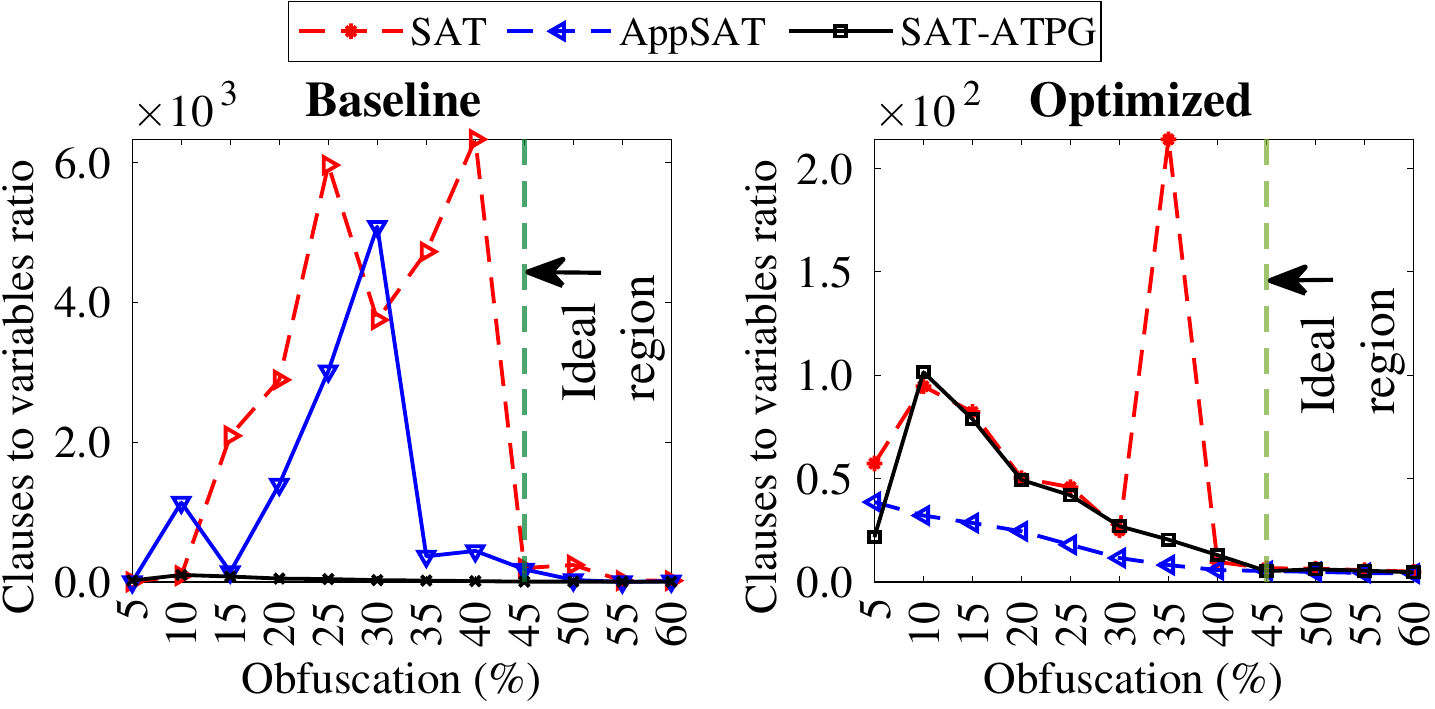}
\caption{The comparison between the baseline and optimized design for the variables to clauses ratio of SAT attacks.}
\centering
\label{fig:sat_ratio}
\vspace{-2mm}
\end{figure}

Additional statistics about the behavior of the SAT solver for the obfuscated \emph{c7552} circuit are given in Fig. \ref{fig:sat_ratio}. The SAT solver determines when a boolean formula is satisfiable or not. One approach to measure the chance of convergence of the attack is to measure the ratio of clauses to variables of the SAT solver. With the help of this ratio, an obfuscated design can be labelled SAT-hard if the ratio is around 4.2 \cite{sat_ratio}. Fig. \ref{fig:sat_ratio} shows the evolution of the number of clauses to variables with respect to the obfuscation level. The x-axis is the obfuscation level (\%) and the y-axis shows the ratio of clauses to variables. We label the right region of Fig. \ref{fig:sat_ratio} as `Ideal region' because the clauses to variables ratio is near the ideal value of 4.2.

Further details are given in Table \ref{tab:obfuscation_ratio} where we also list the key sizes for different designs and two obfuscation rates, namely 55\% and 60\%. Note that, counter-intuitively, the decomposed designs have better variables to clause ratios. Our interpretation is that decomposition keeps keys that are less correlated to one another, thus each individual key bit is more effective. Readers are directed to \cite{sat_ratio} for details on the SAT attack and to \cite{ATPG} for a discussion on key interference.


\begingroup
\setlength{\tabcolsep}{2.0pt} 
\begin{table} [tb]
\footnotesize \centering
\caption{Analysis of variables to clauses ratio for $obf_c=55$\% and 60\%}
\label{tab:obfuscation_ratio}
\begin{tabular}{|p{1.25cm}|p{0.515cm}|p{1.4cm}|p{1.15cm}|p{1.35cm}|p{1.20cm}|p{0.75cm}|}
\hline 
\textbf{Attack} & \textbf{Obf. (\%)} & \textbf{\textit{Key length} \, (bits)} & \textbf{\textit{Variables}} & \textbf{Clauses} & \textbf{Iterations} & \textbf{Ratio} \\ \hline \hline
\multirow{4}{*}{\parbox{1.3cm}{SAT \cite{SAT}}} & 55 & 7494 & 32318070 & 1597559 & 820 & 20.8  \\ \hhline{~------}
& 60 & 8582 & 33315150 & 1898618 & 540 & 17.5 \\ \hhline{~------}
& 55$\ast$ & 4014 & 3434578 & 598599 & 139 & 5.7 \\ \hhline{~------}
& 60$\ast$ & 4406 & 2829008 & 564533 & 106 & 5.0 \\ \hline \hline
\multirow{4}{*}{\parbox{1.3cm}{AppSAT \cite{AppSAT}}} & 55 & 7994 & 15843074 & 1127281 & 8 & 14.0  \\ \hhline{~------}
& 60 & 8582 & 15412584 & 1181330 & 7 & 13.0 \\ \hhline{~------}
& 55$\ast$ & 4014 & 1320652 & 302801 & 2 & 4.36 \\ \hhline{~------}
& 60$\ast$ & 4406 & 1419176 & 346847 & 2 & 4.09 \\ \hline \hline
\multirow{4}{*}{\parbox{1.3cm}{ATPG-SAT \cite{ATPG}}} & 55 & 7494 & 27243806 & 1800999 & 630 & 15.1  \\ \hhline{~------}
& 60 & 8582 & 31166810 & 2247522 & 636 & 13.8 \\ \hhline{~------}
& 55$\ast$ & 4014 & 3642116 & 664817 & 148 & 5.4 \\ \hhline{~------}
& 60$\ast$ & 4406 & 2274084 & 480548 & 85 & 4.7 \\ \hline
\end{tabular}
\footnotesize{$\ast$ Results for the optimized designs}
\vspace{-2mm}
\end{table}
\endgroup

\vspace{-2mm}
\subsection{Oracle-less attacks} \label{subsec:scope}
\noindent \textbf{Goal}: to retrieve a key or a key guess.

\begin{figure}[tb]
\centering 
\includegraphics[width=\linewidth]{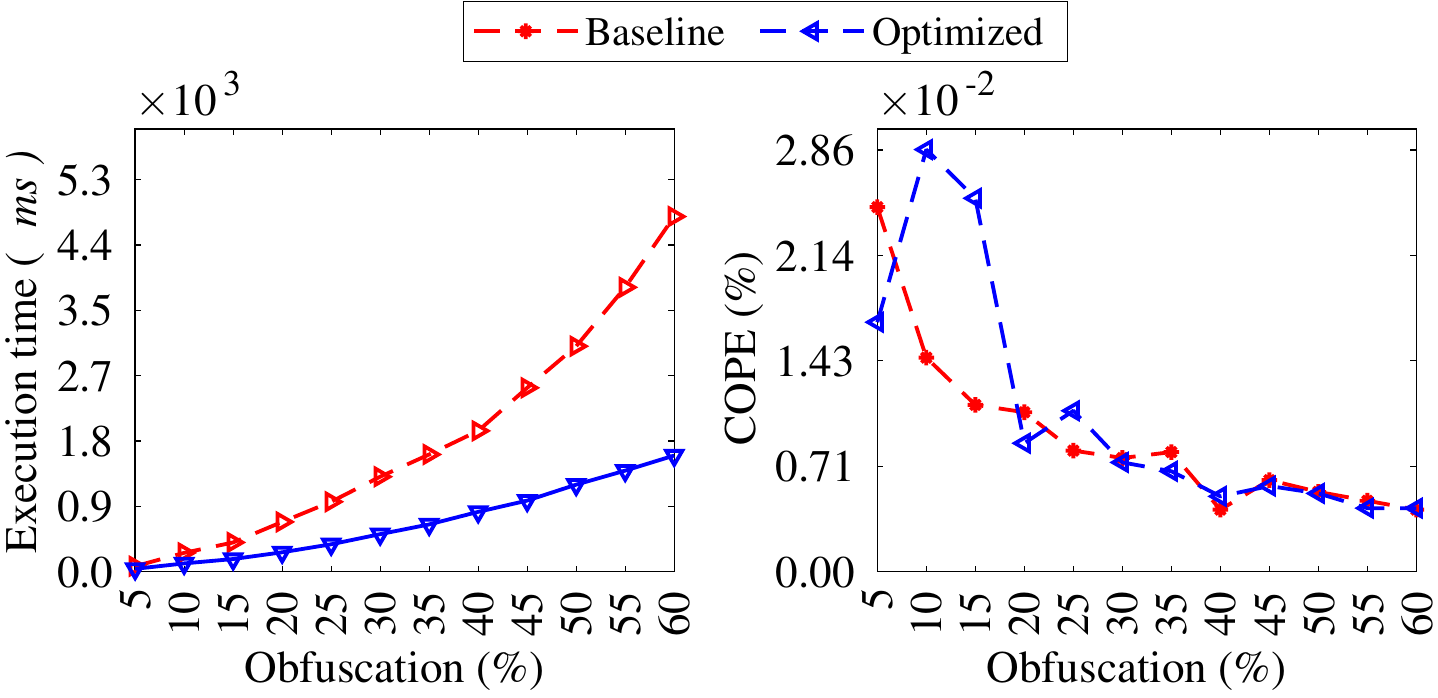}
\caption{The comparison between the baseline and optimized design for the oracle-less SCOPE attack.}
\centering
\label{fig:scope_attack}
\vspace{-2mm}
\end{figure}

Oracle-less attacks do not require an oracle (i.e., a functional IC). Instead, they operate directly on the netlist of the obfuscated circuit. One of such oracle-less attacks is Synthesis-Based Constant Propagation Attack on Logic Locking (SCOPE) \cite{scope}. This attack does not require any knowledge about the locking technique or the obfuscated design. SCOPE performs a synthesis-based analysis on a single key-input port and extracts important design features that may assist to derive the correct key bits. Fig. \ref{fig:scope_attack} illustrates the comparison of execution time, COPE metric for the baseline and optimized design of the \emph{c7552} design. It is clear from the left panel that the execution time is exponentially increasing with the obfuscation level. Similar trends are seem for the baseline and optimized design, both are exponential, but present different rates. The right panel of Fig. \ref{fig:scope_attack} shows the COPE metric, which decreases with the obfuscation level. The details for the calculation of COPE metric are available in \cite{scope}. For simplicity, we clarify that the COPE metric is a rough estimate of the level of vulnerability (\%) to the SCOPE attack.

When SCOPE concludes, a key guess is produced. For each bit of the key, SCOPE assigns either a `1', a `0', or an `X' (undetermined). When matching the guess from SCOPE with our known key, the result is that about 50\% of the key bits are correctly guessed. This percentage is not related to the obfuscation level, the result is always the same for baseline and optimized designs. In other words, SCOPE cannot perform better than a random guess for eASIC. 

\vspace{-2mm}
\section{Discussion} \label{sec:discussion}


Now that we have demonstrated the eASIC concept and its security vs. performance trade-offs, let us discuss other advantages of a hybrid design like our proposed concept. First, just like in FPGAs, the reconfigurability of eASIC allows for design bugs to be fixed by modifying the bitstream, even after the design has been fabricated. From the results, we understand that the obfuscation level should be relatively high to achieve a considerable security, thus the majority of the LUTs should be reconfigurable. This provides an opportunity to correct the issues/bugs that could be easily fixed during the reconfiguration phase. Naturally, there are limitations since a portion of the system is static and cannot be modified. 


It should also be mentioned that eASIC presents a largely regular structure upon visual inspection. This effect can be modulated if it proves to be effective against a reverse engineering adversary. For instance, we could have mapped LUTs of all sizes to LUT\textsubscript{6}, which would increase the layout regularity. Similarly, LUTs could have been laid out in a perfect grid fashion. These two design choices are relatively simple to implement in physical synthesis, but carry overheads that we deemed not advantageous.

A recent trend in obfuscation research is the use of embedded FPGA (eFPGA) \cite{e-FPGA1, e-FPGA2}. While there are advantages to this practice, it has been used selectively to only protect key portions of a design and therefore keep the performance penalty as low as possible. The challenge is in determining which portions/modules of the circuit merit protection and which ones do not. Our eASIC approach bypasses this question almost completely by only revealing (portions of) critical paths when they are selected to become static logic, which we consider an advantage. In \cite{re_2021}, the authors present a top-down methodology to implement ASICs with eFPGAs. Their designs share many of the advantages of our eASIC solution while presenting more regularity than our designs (they make use of logic tiles as in commercial FPGAs). Our tile-free design trades this regularity for performance as evidenced by the layout in Fig. \ref{fig:obfuscation_impl} and the corresponding results in Table \ref{tab:sha_impl}.

\vspace{-2mm}
\section{Conclusions} \label{sec:conclusion}
The main finding of our work is that an eASIC solution contrasts with the current practice of eFPGA obfuscation; our experimental results illustrate that obfuscation rates have to be high to secure the design's intent. To this end, we have presented a custom tool that obfuscates a design and generates an eASIC block. Our LUT decomposition, along with the pin swapping, improves the performance and reduces the area of eASIC designs. We have also validated our results in a commercial physical synthesis tool with industry-strength timing and power analysis. Our security analysis, anchored by the results from diverse attacks, confirms that obfuscation rates should be high. 

As we stated earlier, our goal was to achieve a midpoint design that has the `best of both worlds'; eASIC is precisely that, a solution that combines the characteristics of FPGAs and ASICs into one design. Our future research will focus on other possible benefits of eASIC, including bug fixing, potential side-channel resilience, and further optimizations.

\bibliographystyle{IEEEtran}
\bibliography{obfuscation}


\begin{IEEEbiography}[{\includegraphics[width=1in,height=1.25in,clip,keepaspectratio]{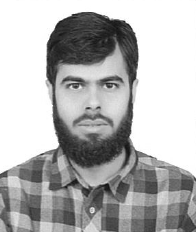}}]{\textbf{ZAIN UL ABIDEEN}} received his M.S. degree in computer engineering (Master in Integration, Security and TRust in Embedded systems) from Grenoble Institute of Technology, Grenoble, France, in 2019. During his master's studies, he was associated with the Cybersecurity Institute Univ. Grenoble Alpes. He worked on hardware security and side channel attacks. He is currently pursuing his doctoral studies at Tallinn University of Technology (TalTech), Tallinn, Estonia. His research work is mainly focused on hardware security and obfuscation-based ASIC design.

\end{IEEEbiography}

\begin{IEEEbiography}[{\includegraphics[width=1in,height=1.25in,clip,keepaspectratio]{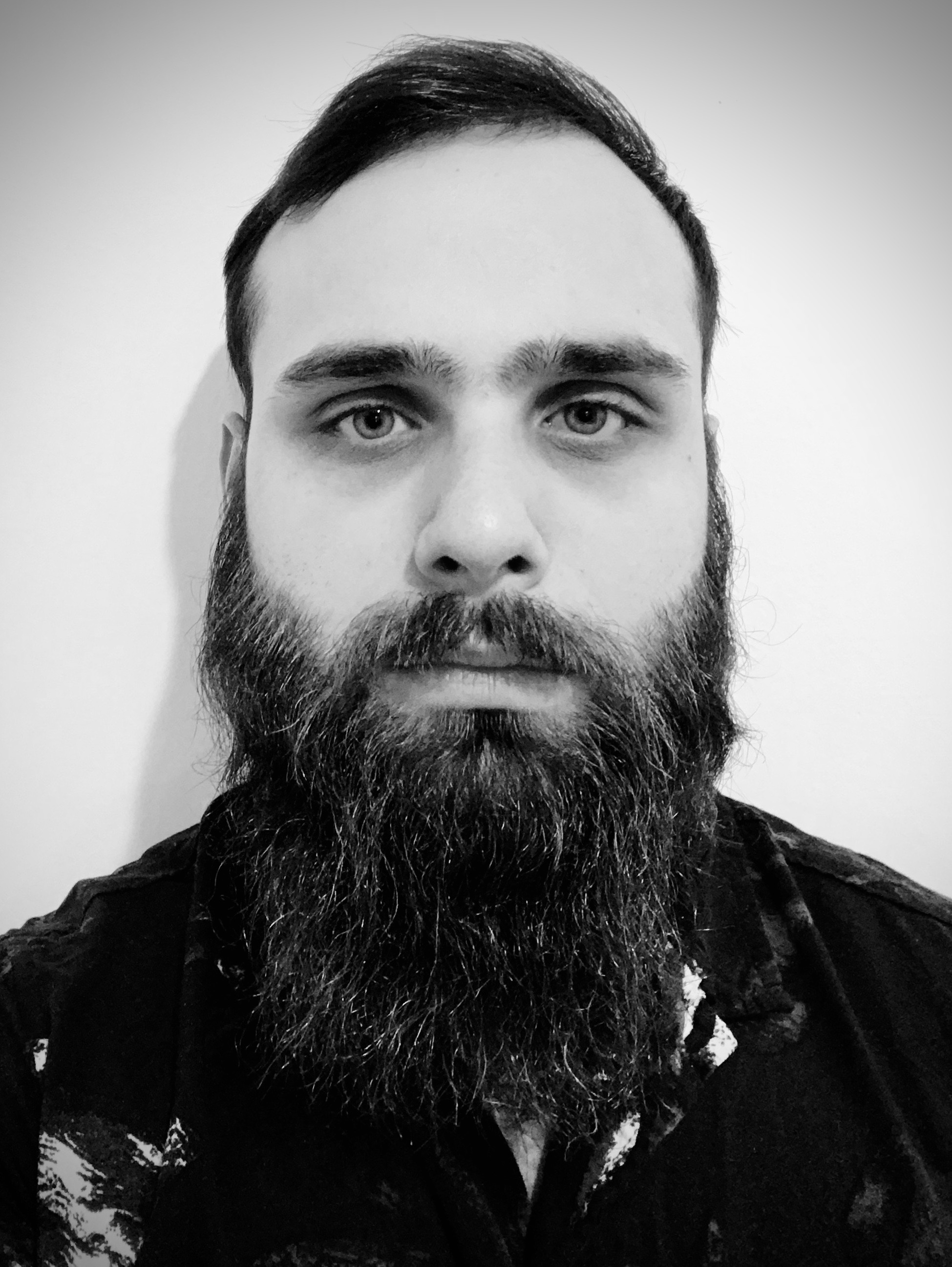}}]{Tiago D. Perez} received the M.S. degree in electric engineering from the University of Campinas, S\~ao Paulo, Brazil, in 2019. He is currently pursuing a Ph.D. degree at Tallinn University of Technology (TalTech), Tallinn, Estonia. From 2014 to 2019, he was a Digital Designer Engineer with Eldorado Research Institute, S\~ao Paulo, Brazil. His fields of work include digital signal processing, telecommunication systems and IC implementation. His current research interests include the study of hardware security from the point of view of digital circuit design and IC implementation.

\end{IEEEbiography}


\begin{IEEEbiography}[{\includegraphics[width=1in,height=1.25in,clip,keepaspectratio]{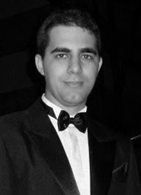}}]{\textbf{Mayler Martins}}
 received the M.S. (summa cum laude) and Ph.D. degrees in microelectronics from the Universidade Federal do Rio Grande do Sul, in 2012, and 2015, respectively. From 2016 to 2018, he was a Research Scientist with the ECE Department, Carnegie Mellon University, Pittsburgh, PA, USA. He worked as a Lead Engineer, from 2018 to 2022, at Siemens EDA, Fremont, USA. His current position is R\&D Engineer Staff at Synopsys, Sunnyvale, CA, USA. His current research interests include logic synthesis methods focusing in QoR optimization.
\end{IEEEbiography}

\begin{IEEEbiography}[{\includegraphics[width=1in,height=1.25in,clip,keepaspectratio]{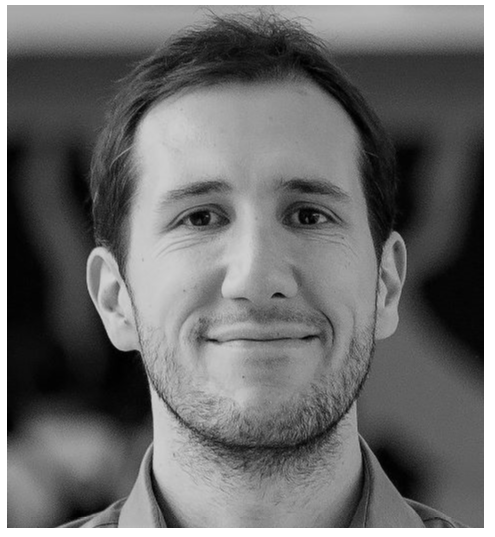}}]{\textbf{Samuel Pagliarini}}
(M'14) received the PhD degree from Telecom ParisTech, Paris, France, in 2013. He has held research positions with the University of Bristol, Bristol, UK, and with Carnegie Mellon University, Pittsburgh, PA, USA. He is currently a Professor with Tallinn University of Technology (TalTech) in Tallinn, Estonia where he leads the Centre for Hardware Security. His current research interests include many facets of digital circuit design, with a focus on circuit reliability, dependability, and hardware trustworthiness.
\end{IEEEbiography}

\end{document}